\documentclass[11pt]{article}
\usepackage{axodraw2,pix}
\usepackage{epsfig}
\usepackage{amsfonts}
\usepackage{amsmath}
\usepackage{amssymb}
\usepackage{bbm,bm}
\usepackage{graphicx}
\usepackage{slashed}
\usepackage{float}
\usepackage[sort&compress]{natbib}
\setcitestyle{numbers,open={[},close={]},comma}
\allowdisplaybreaks[1]

\usepackage{tikz}
\newdimen\nodeDist
\usepackage{datetime}
\usepackage{xcolor}
\usepackage[
  colorlinks=true,
  linkcolor={red!50!black},
  citecolor={blue!50!black},
  filecolor=black,
  urlcolor=black,
  breaklinks=true
  ]{hyperref}

\newcommand{\rmii}[1]{{\mbox{\tiny\rm{#1}}}}
\renewcommand{\vec}[1]{{\bf #1}}
\newcommand{\gammaE}{\gamma_\rmii{E}}
\newcommand{\bmu}{\bar\mu}
\newcommand{\rmO}{{\mathcal{O}}}

\newcommand{\mD}{m_\rmii{D}}

\newcommand{\gE}{g_\rmii{3d}}

\newcommand{\minf}{m_{\infty}}

\newcommand{\QE}{Q_\rmii{E}}
\newcommand{\KE}{K_\rmii{E}}
\newcommand{\PE}{P_\rmii{E}}

\newcommand{\Tint}[1]{{\hbox{$\sum$}\!\!\!\!\!\!\!\int\,}_{\!\!\!\!\raise-0.9ex\hbox{$\scriptstyle{#1}$}}}
\newcommand{\Tinti}[1]{{{\Sigma}\!\!\!\!\raise0.3ex\hbox{$\int$}_\rmii{${#1}$}}}
\newcommand{\Tintip}[1]{{{\Sigma'}\!\!\!\!\!\raise0.3ex\hbox{$\int$}_\rmii{${#1}$}}}

\def\OO{\mathcal{O}}

\def\alphas{\alpha_{\mathrm{s}}}

\def\mDsq{\mD^2}

\def\Tr{\mathrm{Tr}}

\def\Cbp{\mathcal{C}(b_\perp)}

\def\Zg{Z_\mathrm{g}}

\def\Zf{Z_\mathrm{f}}

\def\CR{C_\mathrm{R}}
\def\CA{C_\rmii{A}}
\def\CF{C_\rmii{F}}
\def\CR{C_{\rmii{R}}}
\def\Nc{N_{\rm c}}
\def\dA{d_\rmii{A}}
\def\dR{d_\rmii{R}}

\def\nB{n_{\rmii{B}}}
\def\nF{n_{\rmii{F}}}

\def\Dgtr(#1){D^{>}(#1)}
\def\Dlss(#1){D^{<}(#1)}
\def\DF(#1){D^\mathcal{T}(#1)}
\def\DaF(#1){D^{\overline{\mathcal{T}}}(#1)}

\newcommand{\lambdaE}{\lambda_\rmii{E}}
\def\Nf{N_\rmii{F}}
\def\Tf{T_\rmii{F}}
\def\TR{T_\rmii{R}}
\def\bpp{\vec{p}_{\perp}}
\def\bqp{\vec{q}_{\perp}}

\def\mIsqq{m_{\infty,{\rm q}}^2}
\def\mIsqg{m_{\infty,{\rm g}}^2}

\def\TopoWRF(#1,#2){\;\pic{
  #1(-10,15)(30,15)
  #2(15,15)(15,0,180)%
  \GCirc(0,15){2}{0.75}
  \GCirc(30,15){2}{0.75}
 }}
\def\ToptWBBF(#1,#2,#3){\;\pic{
  #1(-10,15)(30,15)
  #2(15,15)(15,0,180)%
  #3(15,15)(7.5,0,180)%
  \GCirc(0,15){2}{0.75}
  \GCirc(30,15){2}{0.75}
  \Vertex(7.5,15){1}
  \Vertex(22.5,15){1}
 }}
 \def\ToptWSBBF(#1,#2,#3,#4,#5){\;\pic{
  #1(-10,15)(30,15)
  #2(15,15)(15,0,70)
  #3(15,15)(15,110,180)
  \GCirc(15,15 70 sin 15 mul add){5}{0.75}
  #4(15,15 70 sin 15 mul add)(5,0,180)
  #5(15,15 70 sin 15 mul add)(5,180,360)%
  \GCirc(0,15){2}{0.75}
  \GCirc(30,15){2}{0.75}
  }}
\def\ToptWSBSBF(#1,#2,#3){\;\pic{
  #1(-10,15)(30,15)
  #2(6,15)(6,0,180)%
  #3(24,15)(6,0,180)%
  \GCirc(0,15){2}{0.75}
  \GCirc(30,15){2}{0.75}
  \Vertex(12,15){1}
  \Vertex(18,15){1}
  }}
\def\ToptWSBBSF(#1,#2,#3){\;\pic{
  #1(-10,15)(30,15)
  #2(10,15)(10,0,180)%
  #3(20,15)(10,180,360)%
  \GCirc(0,15){2}{0.75}
  \GCirc(30,15){2}{0.75}
  \Vertex(10,15){1}
  \Vertex(20,15){1}
  }}
\def\ToptWSF(#1,#2,#3){\;\pic{
  #1(30,15)(-10,15)
  #2(15,15)(15,0,180)
  #3(15,15)(15,180,360)%
  \GCirc(0,15){2}{0.75}
  \GCirc(30,15){2}{0.75}
  }}
\def\ToptWMF(#1,#2,#3,#4){\;\pic{
  #1(-10,15)(30,15)
  #2(15,15)(15,0,90)
  #3(15,15)(15,90,180)
  #4(15,30)(15,15)%
  \GCirc(0,15){2}{0.75}
  \GCirc(30,15){2}{0.75}
  \Vertex(15,15){1}
  }}
\def\ToptWMnF(#1,#2,#3,#4,#5,#6,#7,#8){\;\pic{
  #1(-10,15)(30,15)
  #2(15,15)(15,0,90)
  #3(15,15)(15,90,180)
  #4(15,30)(15,15)%
  \Textt(0,10,#5)
  \Textt(30,10,#6)
  \Textt(15,10,#7)
  \Textb(15,33,#8)
  \GCirc(0,15){2}{0.75}
  \GCirc(30,15){2}{0.75}
  \Vertex(15,15){1}
  }}
\def\ToptWSalF(#1,#2,#3,#4){\;\pic{
  #1(-10,15)(30,15)
  #2(15,15)(15,0,90)
  #3(15,15)(15,90,180)
  #4(0,30)(15,270,360)
  \GCirc(0,15){2}{0.75}
  \GCirc(30,15){2}{0.75}
  }}
\def\ToptWSalnF(#1,#2,#3,#4,#5,#6,#7){\;\pic{
  #1(-10,15)(30,15)
  #2(15,15)(15,0,90)
  #3(15,15)(15,90,180)
  #4(0,30)(15,270,360)
  \Textt(0,10,#5)
  \Textt(30,10,#6)
  \Textb(15,33,#7)
  \GCirc(0,15){2}{0.75}
  \GCirc(30,15){2}{0.75}
  }}
\def\ToptWalF(#1,#2,#3){\;\pic{
  #1(-10,15)(30,15)
  #2(15,15)(15,0,180)
  #3(7.5,15)(7.5,0,180)%
  \GCirc(0,15){2}{0.75}
  \GCirc(30,15){2}{0.75}
  \Vertex(15,15){1}
  }}
\def\ToptWarF(#1,#2,#3){\;\pic{
  #1(-10,15)(30,15)
  #2(15,15)(15,0,180)
  #3(22.5,15)(7.5,0,180)%
  \GCirc(0,15){2}{0.75}
  \GCirc(30,15){2}{0.75}
  \Vertex(15,15){1}
  }}
\def\ToptWTTF(#1,#2,#3){\;\pic{
  #1(-10,15)(30,15)
  #2(0,22.5)(7.5,-90,270)%
  #3(30,22.5)(7.5,-90,270)%
  \GCirc(0,15){2}{0.75}
  \GCirc(30,15){2}{0.75}
 }}
\def\ToptWTRF(#1,#2,#3){\;\pic{
  #1(-10,15)(30,15)
  #2(0,22.5)(7.5,-90,270)%
  #3(22.5,15)(7.5,0,180)%
  \GCirc(0,15){2}{0.75}
  \GCirc(30,15){2}{0.75}
  \Vertex(15,15){1}
 }}
\def\TopoWR(#1,#2){\;\pic{
  #1(0,15)(30,15)
  #2(15,15)(15,0,180)%
  \GCirc(0,15){2}{0.75}
  \GCirc(30,15){2}{0.75}
 }}
\def\ToptWBB(#1,#2,#3){\;\pic{
  #1(0,15)(30,15)
  #2(15,15)(15,0,180)%
  #3(15,15)(7.5,0,180)%
  \GCirc(0,15){2}{0.75}
  \GCirc(30,15){2}{0.75}
  \Vertex(7.5,15){1}
  \Vertex(22.5,15){1}
 }}
\def\ToptWSBB(#1,#2,#3,#4,#5){\;\pic{
  #1(0,15)(30,15)
  #2(15,15)(15,0,70)
  #3(15,15)(15,110,180)
  \GCirc(15,15 70 sin 15 mul add){5}{0.75}
  #4(15,15 70 sin 15 mul add)(5,0,180)
  #5(15,15 70 sin 15 mul add)(5,180,360)%
  \GCirc(0,15){2}{0.75}
  \GCirc(30,15){2}{0.75}
  }}
\def\ToptWSBSB(#1,#2,#3){\;\pic{
  #1(0,15)(30,15)
  #2(6,15)(6,0,180)%
  #3(24,15)(6,0,180)%
  \GCirc(0,15){2}{0.75}
  \GCirc(30,15){2}{0.75}
  \Vertex(12,15){1}
  \Vertex(18,15){1}
  }}
\def\ToptWSBBS(#1,#2,#3){\;\pic{
  #1(0,15)(30,15)
  #2(10,15)(10,0,180)%
  #3(20,15)(10,180,360)%
  \GCirc(0,15){2}{0.75}
  \GCirc(30,15){2}{0.75}
  \Vertex(10,15){1}
  \Vertex(20,15){1}
  }}
\def\ToptWS(#1,#2,#3){\;\pic{
  #1(30,15)(0,15)
  #2(15,15)(15,0,180)
  #3(15,15)(15,180,360)%
  \GCirc(0,15){2}{0.75}
  \GCirc(30,15){2}{0.75}
  }}
\def\ToptWM(#1,#2,#3,#4){\;\pic{
  #1(0,15)(30,15)
  #2(15,15)(15,0,90)
  #3(15,15)(15,90,180)
  #4(15,30)(15,15)%
  \GCirc(0,15){2}{0.75}
  \GCirc(30,15){2}{0.75}
  \Vertex(15,15){1}
  }}
\def\ToptWMn(#1,#2,#3,#4,#5,#6,#7,#8){\;\pic{
  #1(0,15)(30,15)
  #2(15,15)(15,0,90)
  #3(15,15)(15,90,180)
  #4(15,30)(15,15)%
  \Textt(0,10,#5)
  \Textt(30,10,#6)
  \Textt(15,10,#7)
  \Textb(15,33,#8)
  \GCirc(0,15){2}{0.75}
  \GCirc(30,15){2}{0.75}
  \Vertex(15,15){1}
  }}
\def\ToptWSal(#1,#2,#3,#4){\;\pic{
  #1(0,15)(30,15)
  #2(15,15)(15,0,90)
  #3(15,15)(15,90,180)
  #4(0,30)(15,270,360)
  \GCirc(0,15){2}{0.75}
  \GCirc(30,15){2}{0.75}
  }}
\def\ToptWSaln(#1,#2,#3,#4,#5,#6,#7){\;\pic{
  #1(0,15)(30,15)
  #2(15,15)(15,0,90)
  #3(15,15)(15,90,180)
  #4(0,30)(15,270,360)
  \Textt(0,10,#5)
  \Textt(30,10,#6)
  \Textb(15,33,#7)
  \GCirc(0,15){2}{0.75}
  \GCirc(30,15){2}{0.75}
  }}
\def\ToptWal(#1,#2,#3){\;\pic{
  #1(0,15)(30,15)
  #2(15,15)(15,0,180)
  #3(7.5,15)(7.5,0,180)%
  \GCirc(0,15){2}{0.75}
  \GCirc(30,15){2}{0.75}
  \Vertex(15,15){1}
  }}
\def\ToptWar(#1,#2,#3){\;\pic{
  #1(0,15)(30,15)
  #2(15,15)(15,0,180)
  #3(22.5,15)(7.5,0,180)%
  \GCirc(0,15){2}{0.75}
  \GCirc(30,15){2}{0.75}
  \Vertex(15,15){1}
  }}
\def\ToptWTT(#1,#2,#3){\;\pic{
  #1(0,15)(30,15)
  #2(0,22.5)(7.5,-90,270)%
  #3(30,22.5)(7.5,-90,270)%
  \GCirc(0,15){2}{0.75}
  \GCirc(30,15){2}{0.75}
 }}
\def\ToptWTR(#1,#2,#3){\;\pic{
  #1(0,15)(30,15)
  #2(0,22.5)(7.5,-90,270)%
  #3(22.5,15)(7.5,0,180)%
  \GCirc(0,15){2}{0.75}
  \GCirc(30,15){2}{0.75}
  \Vertex(15,15){1}
 }}

\def\scfc{0.7}  %
\def\phgt{21}   %
\def\pwcb{31.5} %
\def\pwcc{42} %
\newcommand{\PIC}[4]{\;\parbox[c]{#2 pt}{\begin{picture}(#2,#3)(0,0)
\SetWidth{1.0}\SetScale{#4} #1 \end{picture}}\;}
\renewcommand{\picb}[1]{\PIC{#1}{\pwcb}{\phgt}{\scfc}}
\renewcommand{\picc}[1]{\PIC{#1}{\pwcc}{\phgt}{\scfc}}

\makeatletter \@addtoreset{equation}{section} \makeatother
\renewcommand{\theequation}{\arabic{section}.\arabic{equation}}
\makeatletter
\renewcommand\section{\@startsection{section}{1}{\z@}%
  {-5.5ex \@plus -1ex \@minus -.2ex}%
  {2.3ex \@plus.2ex}%
  {\normalfont\large\bfseries}}
\renewcommand\subsection{\@startsection{subsection}{2}{\z@}%
  {-3.25ex\@plus -1ex \@minus -.2ex}%
  {1.5ex \@plus .2ex}%
  {\normalfont\normalsize\bfseries}}
\renewcommand\thesection{\@arabic\c@section}
\renewcommand\thesubsection{\thesection.\@arabic\c@subsection}
\renewcommand{\@seccntformat}[1]{%
  \csname the#1\endcsname.\hspace{1.0em}}
\makeatother
\begin{document}

\flushbottom

\begin{titlepage}

\begin{flushright}
\monthname~\the\year%
\end{flushright}
\begin{centering}

\vfill

{\Large{\bf
The force-force correlator at the hard thermal scale of hot QCD
}}

\vspace{0.8cm}

\renewcommand{\thefootnote}{\fnsymbol{footnote}}
Jacopo~Ghiglieri$^{\rm a,}$%
\footnote{jacopo.ghiglieri@subatech.in2p3.fr},
Philipp~Schicho$^{\rm b,}$%
\footnote{schicho@itp.uni-frankfurt.de},
Niels~Schlusser$^{\rm c,}$%
\footnote{niels.schlusser@unibas.ch},
Eamonn Weitz$^{\rm a,}$%
\footnote{eamonn.weitz@subatech.in2p3.fr}

\vspace{0.8cm}

$^{\rm a}$%
{\em
SUBATECH, Universit\'e de Nantes, IMT Atlantique, IN2P3/CNRS,\\
4 rue Alfred Kastler, La Chantrerie BP 20722, 44307 Nantes, France\\
}

\vspace*{0.3cm}

$^{\rm b}$%
{\em
Institute for Theoretical Physics, Goethe Universit{\"a}t Frankfurt,
60438 Frankfurt, Germany}

\vspace*{0.3cm}

$^{\rm c}$%
{\em
Biozentrum, University of Basel,\\
Spitalstrasse 41,
4056 Basel, Switzerland}

\vspace*{0.8cm}

\mbox{\bf Abstract}

\end{centering}

\vspace*{0.3cm}

\noindent
High-energy particles traversing the Quark-Gluon plasma experience modified (massive) dispersion, although their vacuum mass is negligible compared to the kinetic energy. 
Due to poor convergence of the perturbative series in the regime of soft loop momenta, 
a more precise determination of this effective mass is needed.
This paper continues our investigation on the factorisation between
strongly-coupled infrared classical and
perturbative ultraviolet behavior. 
The former has been studied non-perturbatively within EQCD by determining a non-local operator on the lattice.
By computing the temperature-scale contribution to the same operator in 4D QCD
at next-to-leading order (NLO),
we remove the ultraviolet divergence of the EQCD calculation with an opposite infrared divergence from
the hard thermal scale.
The result is a consistent, regulator-independent determination of the classical contribution
where the emergence of new divergences signals sensitivities to new regions of phase space.
We address the numerical impact of the classical and NLO thermal corrections on the convergence of
the factorised approach and on
the partial applicability of our results to calculations of transport coefficients.

\vfill

\vfill
\end{titlepage}

{\hypersetup{hidelinks}
\tableofcontents
}
\clearpage

\section{Introduction}
\la{sec:introduction}
The Quark-Gluon Plasma (QGP) is a deconfined, chirally symmetric 
phase of nuclear matter as described by Quantum Chromodynamics (QCD). The QGP is actively investigated 
through heavy-ion collisions. Disentangling the properties 
of this short-lived state from the shower of particles in 
the detectors requires an interplay of observables in the two main 
classes of \textit{bulk properties} and \textit{hard probes}. The former pertain
to the collective properties of the many lower-energy particles, while
the latter refer to the few energetic and/or feebly-coupled ones 
which are not expected to equilibrate. 

Jets are one of the most important hard probes --- see \cite{Connors:2017ptx,Cunqueiro:2021wls,Apolinario:2022vzg} 
for recent reviews.
These self-collimated showers of hadrons are seeded by 
highly-energetic partons, in a regime where perturbative QCD should be
applicable. Whether that remains true as these partons cross the QGP is still an open question, thus influencing
the methods that one should apply to describe jets in medium.
If the jet and the medium are weakly coupled --- which does not 
necessarily imply a weakly-coupled medium --- then the main theory 
ingredient in jet modification in the QGP is
{\em medium-induced radiation}.
These modifications are
sourced by the jet partons undergoing 
frequent exchanges of transverse 
momentum with the medium through elastic scatterings, encoded in the 
transverse scattering kernel $\Cbp$ . The jet
partons can also undergo forward scattering, whereby they 
are first scattered off their original momentum state and then back into 
it, thus shifting the dispersion relation by an amount 
named $\minf$, the {\em asymptotic mass}. 

The transverse scattering kernel $\Cbp$ and $\minf$ are
the most important medium-dependent ingredients in the description
of medium-induced radiation, see e.g.~\cite{Arnold:2002ja}.
In this paper, we continue our investigations of $\minf$, 
started in \cite{Moore:2020wvy,Ghiglieri:2021bom}. 
We recall that  these mass shifts are written in terms of
the gauge condensate $\Zg$ and
the fermion condensate $\Zf$ \cite{CaronHuot:2008uw}
\begin{equation}
\label{eff_mass}
\mIsqq = g^2\CF^{ }\left( \Zg^{ } + \Zf^{ } \right)
\,, \qquad
\mIsqg = g^2\CA^{ }\Zg^{ } + 2g^2 \Tf^{ }\Nf^{ }\Zf^{ }
\,,
\end{equation}
where
$\mIsqq$ applies for quarks and
$\mIsqg$ applies for gluons.
Here
$\CF=(\Nc^2-1)/(2\Nc)$ is the quadratic Casimir of the quark representation,
$\CA=\Nc$ is the adjoint Casimir,
$\Nf$ is the number of light (Dirac) quark species, and
$\Tf=\frac{1}{2}$.

Eq.~\eqref{eff_mass} can be understood as arising from
integrating out the energy scale of the jet $E\gg T$ and
truncating at first order in $T/E$. 
The matching coefficients are determined at first order in $g$ ---
we will return to this issue later.
If the scale of the hard parton is $E\gtrsim T$ rather than $E\gg T$,
higher orders in the $T/E$ expansion become relevant. This also makes 
clear that ours is a distinct 
problem from the one recently tackled in 
\cite{Gorda:2022fci,Ekstedt:2023anj,Ekstedt:2023oqb,Gorda:2023zwy}, namely the 
two-loop and power corrections to Hard Thermal Loops and their
relation to the asymptotic mass for quasiparticles obeying $T\gg p\gg g T$.

The condensates $\Zg$ and $\Zf$
are non-local and
have a gauge-invariant definition %
in terms of the correlators
in the Hard Thermal Loop (HTL) action~\cite{Braaten:1991gm,CaronHuot:2008uw}
\begin{align}
\label{mom_ferm_cond}
  \Zf &\equiv \frac{1}{2\dR} \Big\langle
    \overline{\psi} \frac{v_\mu\gamma^\mu}{v\cdot D} \psi
  \Big\rangle
  \;,\\
\label{mom_gauge_cond}
  \Zg &\equiv  \frac{1}{\dA} \Big\langle
    v_{\alpha}^{ } F^{\alpha\mu}
    \frac{1}{(v \cdot D)^2}
    v_{\nu}^{ } F^\nu_{\;\; \mu}
  \Big\rangle
  \;,
\end{align}
where $v^{\mu}=(1,\vec{v})$ is the light-like four-velocity of the hard particle,
$d_{\rmii{R}}$ is the dimension of the fermion,
$d_{\rmii{A}}$ the dimension of the adjoint representation,
and the expectation value $\langle\dots\rangle$ denotes a thermal expectation value.
Appendix~\ref{sec_conv} specifies our conventions.

All operators in thermal QCD
are sensitive not just to the contribution of thermal modes, with 
momenta of the order of the temperature $T$. There exist also 
\emph{soft modes}, with momenta  $p\sim gT$, and gluonic ultrasoft 
modes, with momenta of $\mathcal{O}(g^2T)$.
The former cause the perturbative 
expansion to be in $g$ rather than $\alphas$ and are largely responsible
for its slow convergence. The latter,
at an operator-dependent order, cause the loop expansion to 
fail~\cite{Linde:1980ts}. Moreover, for bosons these infrared (IR) modes
are dominated by classical field dynamics, as their occupation numbers
are $\nB(p)\approx T/p\gg 1$, with $\nB$ the Bose--Einstein distribution.

Soft modes first contribute to $\Zg$ at relative $\mathcal{O}(g)$, whereas
$\Zf$ is first expected to receive soft corrections at higher orders~\cite{CaronHuot:2008uw},
i.e.
\begin{equation}
  \Zg^{ }
  = \Zg^\rmii{LO} + \delta\Zg^{ }
    = \frac{T^{2}}{6} - \frac{T \mD}{2\pi}
    + \rmO(g^2)
  \,,\qquad
  \Zf^{ }
    = \Zf^\rmii{LO} + \delta\Zf^{ }
    = \frac{T^{2}}{12} + 0
    + \rmO(g^2)
  \,,
  \label{nlocond}
\end{equation}
where $\mD^{2}$ is the leading-order Debye mass, 
given by~\cite{Weldon:1982aq}
\begin{equation}
  \mDsq =
    \frac{g^2 T^2}{3} \left( \CA + \Tf \Nf \right)
    + \OO(g^4)
  \;.
\end{equation}
The next-to-leading order (NLO) correction,
$\delta \Zg$, is negative and, for values of the coupling corresponding to QGP 
temperatures, of comparable magnitude to the leading-order (LO) contribution $\Zg^\rmii{LO}$.
The main motivation for \cite{Moore:2020wvy,Ghiglieri:2021bom} and 
for the present paper is then  to use an interplay of lattice and 
perturbation theory to  determine the classical IR contribution to all orders.
This can also allow to extend the applicability of the evaluation of the 
mass shift to lower temperatures, since it uses perturbation theory for the 
thermal modes only, which still have $\alphas$ as expansion parameters.

At the technical level, this has been made possible by developments 
introduced by Caron-Huot in \cite{CaronHuot:2008ni,CaronHuot:2008uw}.
These papers showed how, for a hard, light-like parton interacting with 
the QGP, this classical contribution can be treated in Electrostatic QCD
(EQCD)~\cite{Braaten:1994na,Braaten:1995cm,Braaten:1995jr,Kajantie:1995dw,Kajantie:1997tt},
a dimensionally-reduced Effective Field Theory (EFT) of QCD.
This allows for
non-perturbative evaluations of the EQCD contribution to $\Cbp$ and $\Zg$ through lattice EQCD.
We refer to 
\cite{Moore:2019lgw,Moore:2021jwe,Schlichting:2021idr} for the 
evaluation of $\Cbp$ in EQCD, its merging with the $T$-scale contribution
and its impact on medium-induced radiation.

For $\minf$, one can rely on the equivalence between inverse covariant
derivatives and integrals over Wilson lines to rewrite Eq.~\eqref{mom_gauge_cond} as an integral
over the light-like separation of the two field strengths, inserted 
over light-like Wilson lines --- see Eq.~\eqref{eq:zg_wline}. 
In \cite{Moore:2020wvy,Ghiglieri:2021bom} its EQCD counterpart was measured
on the lattice as a function of the separation.
To perform the separation integration, Ref.~\cite{Ghiglieri:2021bom} also provided a NLO perturbative 
determination in EQCD, to be used at distances shorter than the lattice 
spacing.
There, in addition to discretisation, one faces another conceptual obstacle:
the ultraviolet (UV) of EQCD is not, by construction, the UV of 4D thermal QCD. This is 
made manifest by the emergence of power-law and log-divergences from
the integration of the LO (one-loop) and NLO (two-loop) perturbative EQCD at short separations.
The super-renormalizable nature of EQCD prevents further UV divergences at higher orders.

In \cite{Ghiglieri:2021bom} the separation integral could then be carried out
by subtracting the LO and NLO UV-divergent behaviors. For the power-law 
divergent LO term, this only amounts to avoiding double counting with the 
soft limit of the $T$-scale LO contribution $\Zg^\rmii{LO}$.
For the log-divergent NLO term, this subtraction introduces a
cutoff scale $L_\rmi{min}$. Physically, it signals an expected 
IR logarithmic divergence from thermal modes in 4D QCD at two-loop order.

The goal of this paper is precisely to determine the contribution 
to $\Zg$ from thermal modes in QCD at two-loop level.
In so doing, we 
will recover the expected IR divergence, evaluate it in 
\emph{dimensional regularisation} (DR) together with the 
aforementioned subtraction, thus lifting the dependence 
on the cutoff scale $L_\rmi{min}$ and providing 
an expression for $\Zg$ that is two-loop accurate for the $T$-scale modes
and, when combined with the lattice results from \cite{Ghiglieri:2021bom},
all-order accurate for the classical IR modes.
We anticipate that we shall also find new ultraviolet and collinear 
divergences, signaling the emergence, at this loop order, of new regions of 
phase space and a potential sensitivity to one-loop
corrections to the matching coefficients in Eq.~\eqref{eff_mass}.
Part of our main results have been anticipated in \cite{Ghiglieri:2023cyw,eamonnthesis}.
We also address which part of our results here and in 
\cite{Moore:2020wvy,Ghiglieri:2021bom} remains valid 
when the energy of the parton becomes of the order of the temperature,
which is the relevant regime for calculations of transport coefficients
and production rates.

The paper is organised as follows.
Section~\ref{sec:background} summarises
the setup of the problem, the EQCD evaluation and the subtraction 
procedure for the UV divergences introduced in \cite{Ghiglieri:2021bom}.
In Sec.~\ref{sec:O:gsq:calculation} we present our two-loop calculation of the $T$-scale
contribution in 4D QCD, showing how, in addition to expected IR  and vacuum UV 
divergences, we find extra ones and discuss their origin.
In Sec.~\ref{sec:matching:to:EQCD},
we show how the IR divergences cancel against the UV ones, leading to a finite,
regulator-independent result for the classical, non-perturbative contribution.
In Sec.~\ref{sec:discussion}, we summarise our main findings and  draw our conclusions.
The appendices are dedicated to the conventions and technical details of our calculations.

\section{Background}
\la{sec:background}
As we anticipated, $\Zg$ can be rewritten as the following integral over 
light-like separated field-strength insertions on a light-like Wilson line
\begin{equation}
\label{eq:zg_wline}
    \Zg=\frac{1}{\dR \CR}\int_{0}^{\infty}\!{\rm d}x^+\,x^+\Tr \Bigl\langle
      U_\rmii{R}(-\infty;x^+)v_{\mu}F^{\mu\nu}(x^+)
      U_\rmii{R}(x^+;0)v_{\rho}F^{\rho}_{\nu}(0)
      U_\rmii{R}(0;\infty)
  \Bigr\rangle
  \,,
\end{equation}
where the Wilson line itself reads
\begin{equation}
\label{eq:4d:Wilson:line:expl}
  U_\rmii{R} (x^+;0) = \mathrm{P} \; \exp \biggl( i g \int_0^{x^+} \!{\rm d} y^+ A^{-a}(y^+) \TR^a \biggr)
  \,,
\end{equation}
after having fixed $v^\mu=(1,0,0,1)$ --- see Appendix~\ref{sec_conv}.
The Wilson lines stretch from and back to negative light-cone 
infinity.
They descend from the relation to the density matrix
of the hard parton, which is defined in the asymptotic past.
\begin{figure}[t]
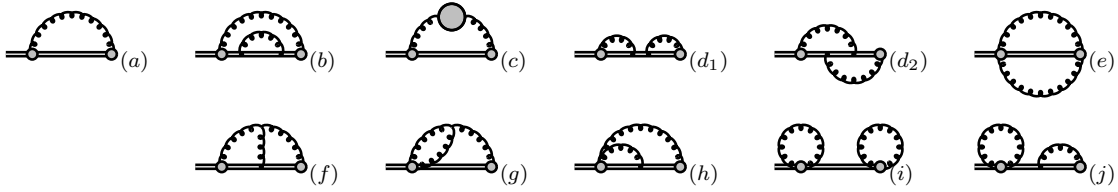

  \centering
  \begin{align*}
    \TopoWRF(\Lsri,\Agliii)_{(a)}\;
    && &
    \ToptWBBF(\Lsri,\Agliii,\Agliii)_{(b)}\;
    &&
    \ToptWSBBF(\Lsri,\Agliii,\Agliii,\Asai,\Asai)_{(c)}\;
    &&
    \ToptWSBSBF(\Lsri,\Agliii,\Agliii)_{(d_1)}\;
    &&
    \ToptWSBBSF(\Lsri,\Agliii,\Agliii)_{(d_2)}\;
    &&
    \ToptWSF(\Lsri,\Agliii,\Agliii)_{(e)}\;
    \nn[3mm]
    && &
    \ToptWMF(\Lsri,\Agliii,\Agliii,\Lgliii)_{(f)}\;
    &&
    \ToptWSalF(\Lsri,\Agliii,\Agliii,\Agliii)_{(g)}\;
    &&
    \ToptWalF(\Lsri,\Agliii,\Agliii)_{(h)}\;
    &&
    \ToptWTTF(\Lsri,\Agliii,\Agliii)_{(i)}\;
    &&
    \ToptWTRF(\Lsri,\Agliii,\Agliii)_{(j)}\;
  \end{align*}
  \caption{%
    Diagrams contributing
    the QCD force-force correlator $\Zg$~\eqref{eq:zg_wline} at
    leading and
    next-to-leading order.
    An external gray shaded vertex denotes an
    $F^{-\perp}$ insertion;
    internal 2-point blobs the respective self-energy;
    a solid line a Wilson line; and
    a curly line a gauge boson ($A_\mu$).
  }
  \label{fig:diagrams}
\end{figure}

At leading order,
the Wilson lines source no gluons and one just connects the two
field-strength tensors with a free thermal gluon propagator,
as in diagram $(a)$ of Fig.~\ref{fig:diagrams}.
It yields \cite{Klimov:1981ka,Klimov:1982bv,Weldon:1982bn}
\begin{eqnarray}
    \label{zglo}
    \Zg^\rmii{LO}&=&
    (D-2)\int_Q[\theta(q^0)+\nB (\vert q^0\vert)] 2\pi\delta(Q^2)=
    (D-2)\int_Q\nB (\vert q^0\vert)2\pi\delta(Q^2)
    \nn
    &\stackrel{D = 4}{=}&
    2\int_{\vec{q}} \frac{\nB(q)}{q}=\frac{T^2}{6}
    \,,
\end{eqnarray}
where
$\nB(q^0)\equiv(\exp(q^0/T)-1)^{-1}$ is
the Bose--Einstein distribution. 
Dimensional regularisation in $D=4-2\epsilon$ ($d=3-2\epsilon$ for the spatial dimensions)
has been used to remove the scale-free vacuum contribution proportional to $\theta(q^0)$.
See Appendix~\ref{sec_conv} and Eq.~\eqref{def_integrals}
for our conventions.

\subsection{Power counting and hierarchy of scales in jet-medium interactions}
\la{subsec:power:counting}

The LO contribution to $\Zg$ arises from the scale $T$ as seen
in Eq.~\eqref{zglo}.
In the infrared $\nB(q)\approx T/q$, so that Eq.~\eqref{zglo} has an order-$g$ sensitivity to
the soft scale $gT$, {\em viz.}\ $\int_{\vec{q}\sim gT} \frac{T}{q^2}\sim gT^2$. 
Indeed, that is where the NLO correction in Eq.~\eqref{nlocond} appears.
We can then summarise the perturbative expansion of $\Zg$  as follows \cite{Ghiglieri:2021bom}
\begin{align}
  &
  \hphantom{\bigg[\quad}
  \mathrm{scale\;}T
  &&
  \mathrm{scale\;}gT
  &&
  \mathrm{scale\;}g^2T
  &
  \nn
  \Zg =
    &\bigg[\quad
      \frac{T^2}{6}
    - \frac{T\mu_\rmi{h}^{ }}{\pi^2}
    &&
    &&
    &\bigg]
  \label{loline}\\
  + &\bigg[\quad
    &&
    - \frac{T\mD}{2\pi}
    + \frac{T\mu_\rmi{h}^{ }}{\pi^2}
    &&
  &\bigg]
  \label{nloline}\\
  + &\bigg[\quad
      c^{\ln}_{\rmi{hard}} \ln\frac{T}{\mu_\rmi{h}}
    + c_{\rmii{$T$}}^{ }
    &&
    + c^{\ln}_{\rmi{hard}} \ln\frac{\mu_\rmi{h}}{\mD} 
    + c^{\ln}_{\rmi{soft}} \ln\frac{\mD}{\mu_\rmi{s}}
    + c_{\rmii{$gT$}}^{ }
    &&
    + c^{\ln}_{\rmi{soft}} \ln\frac{\mu_\rmi{s}}{g^2T}
    + c_{\rmii{$gT^2$}}^{ }
  &\bigg]
  \label{nnloline}\\[2mm]
  +&\,\OO(g^3)\;.&
  &&
\end{align} 
We highlight the contributions from different orders of the coupling $g$ row by row 
and from different scales column by column. 
The  $T$-scale contribution at $\OO(g^0)$ from Eq.~\eqref{zglo} is displayed in \eqref{loline}
and the first $\OO(g)$ correction from soft modes in \eqref{nloline}. 
At the next order, all scales contribute: the completely non-perturbative color-magnetic 
scale $g^2 T$, as well as the soft modes of the color-electric scale $g T$, and the thermal modes 
from $T$. 
The coefficients $c^{\ln}_{\rmi{hard}}$ and $c^{\ln}_{\rmi{soft}}$ were determined 
as a byproduct of \cite{Ghiglieri:2021bom}, whose main aim was the non-perturbative,
all-order determination of the classical soft and ultrasoft contribution.
As we anticipated, the procedure used there introduced an artificial 
dependence on a cutoff $L_\mathrm{min}$: for the moment we have kept the 
notation in Eqs.~\eqref{loline}--\eqref{nnloline} generic,
with $\mu_\rmi{h}$ and $\mu_\rmi{s}$ taken as placeholders for an arbitrary
scheme used to separate the thermal, soft and ultrasoft scales. The coefficients 
on Eq.~\eqref{nnloline} are then to be understood as \emph{scheme-depedent}. For reasons that will become evident later, knowing all coefficients in \eqref{loline}, \eqref{nloline}, and \eqref{nnloline} does not yield a self-contained estimate of $\minf$ at $\OO(g^2)$, yet.

Consequently, our main aim is to %
connect the scheme used to remove the unphysical 
EQCD UV from the EQCD results in \cite{Ghiglieri:2021bom}, giving rise to 
$L_\mathrm{min}$ dependence, to dimensional regularisation. We will 
also compute the $T$-scale contribution in Eq.~\eqref{nnloline}, $c^{\ln}_{\rmi{hard}}$
and $c_{\rmii{$T$}}^{ }$, in that same scheme.
To this end,
we first briefly review the determination of
the EQCD contribution to $\Zg$ from~\cite{Moore:2020wvy,Ghiglieri:2021bom}.

\subsection{EQCD setup}
\label{subsec:EQCD}

One starts from the EQCD continuum action,
\begin{equation}
\label{eq:EQCD_cont_action}
S_{\rmii{EQCD}} = \int_{\vec{x}}\bigg\{
    \frac{1}{2} \Tr\,F_{ij} F_{ij}
  + \Tr\,[D_i, \Phi] [D_i, \Phi]
  + \mDsq \Tr\,\Phi^2
  + \lambdaE^{}(\Tr\,\Phi^2)^2
\bigg\}
  \, ,
\end{equation}
where $\Phi$ is the adjoint, massive scalar arising from dimensional reduction of the $A^0$ gauge field.
One then derives the non-unitary EQCD analog of Eq.~\eqref{eq:zg_wline}
\begin{equation}
\label{eq:wline_eqcd} 
    \tilde{U}_\rmii{R}(L;0)=\mathrm{P}\; \exp \biggl( i \gE \int_0^{L} \!{\rm d}z\left(A_z^a(z)+i\Phi^a(z)\right)  \TR^a \biggr)
    \,.
\end{equation}
Since the field operators are classical in EQCD, they commute and allow us to write
\begin{equation}
  \frac{1}{\dR \CR}\Tr\langle
    \tilde{U}_\rmii{R}(-\infty;L)\mathcal{O}^{a}(L)\TR^{a}
    \tilde{U}_\rmii{R}(L;0)\mathcal{O}^{b}(0)\TR^{b}
    \tilde{U}_\rmii{R}(0;-\infty)\rangle
    =\frac{\TR}{\dR\CR}\langle\mathcal{O}^a(L)\tilde{U}^{ab}_{\rmii{A}}(L;0)\mathcal{O}^{b}(0)\rangle
    \,.
\end{equation}
Thus, the EQCD equivalent of $\Zg$ reads, given rotational invariance in the transverse plane 
\begin{equation}
    Z_{\mathrm{g}}^{3\text{d}}=-\frac{(d-1) T}{\dA}\int_{0}^{\infty}\!{\rm d}L\,L \Bigl\langle\Big(F^{a}_{xz}(L)+i(D_x\Phi(L))^a\Big)
    \tilde{U}^{ab}_{\rmii{A}}(L;0)\Big(F^{b}_{xz}(0)+i(D_x\Phi(0))^b\Big)\Bigr\rangle
    \,,
\end{equation}
where we used $\dR\CR/\TR=\dA$.
At LO in perturbative EQCD,
we can connect the operators with EQCD propagators to find~\cite{CaronHuot:2008uw}
\begin{equation}
    Z_{\mathrm{g}\,\rmii{LO}}^{3\text{d}}=T\int_{\vec{q}} \frac{1}{(q^z+i\varepsilon)^2}
    \bigg[-\frac{q_\perp^2}{q^2+\mD^2}+
    \frac{q_\perp^2}{q^2}+ \frac{(d-1)q_z^2}{q^2}\bigg]=-\frac{T\mD}{2\pi}
    \,,
    \label{loeqcd}
\end{equation}
where the massive propagator is the $\Phi$ propagator. Here we used 
DR to get to a finite result: in so doing, a linearly-UV divergent term gets discarded.
It 
is instructive to note that this term would be $(d-1)\int_{\vec{q}}\frac{T}{q^2}$, which corresponds 
precisely to the IR sensitivity of Eq.~\eqref{zglo}, as argued in the previous subsection, and 
to the linear terms in $\mu_\rmi{h}$ in Eqs.~\eqref{loline} and \eqref{nloline}.
Thus, dimensional regularisation corresponds to subtracting off that $(d-1)\int_{\vec{q}}\frac{T}{q^2}$ 
bare, soft limit from the resummed result in Eq.~\eqref{loeqcd}.
As we shall now see, this does not happen automatically
in the scheme that follows from the lattice computation.

For that end, it is 
more convenient to set $d=3$ and to work with the quantities
\begin{align}
    \label{def:cond:EE}
      \langle EE\rangle &\equiv \frac{1}{2}
        \bigl\langle (D_x \Phi(L))^a\,\tilde{U}^{ab}_{\rmii{A}}(L,0)\,(D_x\Phi(0))^b \bigr\rangle
        \;,\\
    \label{def:cond:BB}
      \langle BB\rangle &\equiv \frac{1}{2} \bigl\langle F_{xz}^a(L)\,\tilde{U}^{ab}_{\rmii{A}}(L;0) \, F_{xz}^b(0) \bigr\rangle
        \;,\\
    \label{def:cond:EB}
      i\langle EB\rangle &\equiv
          \frac{i}{2} \bigl\langle (D_x \Phi(L))^a\,\tilde{U}_{\rmii{A}}^{ab}(L;0) \, F_{xz}^b(0) \bigr\rangle
       + \frac{i}{2} \bigl\langle F_{xz}^a(L) \, \tilde{U}_{\rmii{A}}^{ab}(L;0) \, (D_x \Phi(0))^b \bigr\rangle
      \;,
\end{align}
which leaves us with 
\begin{align}
    Z_{\mathrm{g}}^{3\text{d}}&=-\frac{4 T}{\dA}\int_{0}^{\infty}\!{\rm d}L\,L \Big(-\langle EE\rangle +\langle BB\rangle+i\langle EB\rangle  \Big)%
   =\frac{4 T}{\dA}\int_{0}^{\infty}\!{\rm d}L\,L\,\langle FF\rangle \label{eq:zg3d}
   \,,
\end{align}
where we have implicitly defined
$\langle FF\rangle\equiv \langle EE\rangle -\langle BB\rangle-i\langle EB\rangle$. 

In \cite{Moore:2020wvy,Ghiglieri:2021bom} continuum-extrapolated lattice EQCD determinations of the three 
components of $\langle FF\rangle$ have been performed for distances $0.25<\gE^2 L<3.0$
at four values of the temperature.
To integrate over all separations, this
lattice determination must be complemented at both long and short distances to provide a
finite result for the all-order classical contributions from the soft and ultrasoft scales.
In more detail, the IR region is addressed with a fitting ansatz, intended to reproduce the
expected exponential falloff from electrostatic and magnetostatic screening, since one cannot 
reach arbitrarily large distances on the lattice. As for the UV,
the lattice approach becomes impractical at short distances due to associated discretisation effects.
However, at the shortest available separations, where $\mD L\ll 1$, the lattice determination agrees well with the perturbative NLO EQCD %
determination, as expected. This then suggests a switch to the perturbative 
EQCD determination at short distances. Yet, the naive integration down to zero separation 
of the perturbative EQCD result would result in power-law and log divergences, as we mentioned.

But how does this divergent behavior manifest itself? Because 
of the super - renormalizability of EQCD, in the UV, it is natural to expect from dimensional analysis that
\begin{equation}
    \langle FF\rangle_{m_{D}L\ll 1}=\frac{c_0}{L^3}+\frac{c_2\gE^2}{L^2}+\frac{c_4\gE^4}{L}+...\,.
\end{equation}
The first and second terms above are divergent when inserted into Eq.~\eqref{eq:zg3d}. 
If integrated down to a $L_\mathrm{min}>0$, the former gives rise to a $c_0/L_\mathrm{min}$
contribution. In this position-space scheme, this corresponds to the linear term 
in $\mu_\rmi{h}$ in Eq.~\eqref{nloline}. As the previous discussion should have made clear,
such a term corresponds to the soft limit of the bare LO calculation in Eq.~\eqref{zglo};
as it is already included there, it should be subtracted off from the EQCD calculation,
as per \cite{Moore:2020wvy}.

The coefficient
$c_2$ was computed in~\cite{Ghiglieri:2021bom} and, being a logarithmic divergence,
shows up also in DR.
To arrive at a determination for $Z_{\mathrm{g}}^{\rmi{3d}}$,
\cite{Ghiglieri:2021bom} merged together the perturbative, intermediate and IR regions as follows
\begin{align}
  Z_{\mathrm{g}}^{\rmi{3d}}\big\vert^\text{merge} =\frac{4T}{\dA}\bigg\{&
    \int_0^{L_\text{min}} \!{\rm d}L\,L\bigg[
        \langle FF\rangle_\rmii{NLO}
      - \frac{c_0}{L^3}
      - \frac{c_2 \gE^2}{L^2}\bigg]
  + \int_{L_\text{min}}^{L_\text{max}} \!{\rm d}L\,L\bigg[
      \langle FF\rangle_\text{lat}
      -\frac{c_0}{L^3}\bigg]
    \nn +&
    \int_{L_\text{max}}^\infty \!{\rm d}L\,L\bigg[
        \langle FF\rangle_\text{tail}
      - \frac{c_0}{L^3}\bigg]\bigg\}
    \,,
    \label{eq_subtr}
\end{align}
where \cite{Moore:2020wvy,Ghiglieri:2021bom}
\begin{equation}
    \label{subtrterms}
    c_0=\frac{\dA }{4\pi}\,,\qquad
    c_2=\frac{\CA\dA}{(4\pi)^2}
    \,.
\end{equation}
The shortest and longest
separations for which lattice data is available are
$L_{\text{min}} = 0.25/\gE^2$ and
$L_{\text{max}} = 3.0/\gE^2$.%
\footnote{%
The $\langle EB\rangle$ contribution was only available for $L>0.5/\gE^2$ in \cite{Ghiglieri:2021bom}. 
This is irrelevant 
from the standpoint of $c_2$ subtraction, since only $\langle EE\rangle$ is
logarithmically divergent at small $L$.
}
Between these two separations, the lattice-determined $\langle FF\rangle_\text{lat}$ is used. At asymptotically
large distances the modeled IR tail $\langle FF\rangle_\text{tail}$ is used and at short distances 
the NLO perturbative EQCD result $ \langle FF\rangle_\text{NLO}$ is integrated. As we mentioned,
the $c_0$ term must be subtracted off everywhere, whereas the
$c_2$-term is only subtracted in the perturbative region.
Eq.~\eqref{eq_subtr}
thus introduces an
artificial logarithmic dependence on its boundary $L_\mathrm{min}$.

In Sec.~\ref{sec:matching:to:EQCD}
we will show how this subtraction can be recast in dimensional regularisation.
The resulting UV pole cancels against an opposite IR one from 
 computing the two-loop contribution to $\Zg$ in thermal 4D QCD,
which shall be carried out in the next Section.
This is then our main motivation, as their 
 sum provides a finite result for the non-perturbative contribution, with the artificial cutoff, $L_{\text{min}}$, lifted.
Such a calculation is evidently also in line with our ultimate goal of determining all $\mathcal{O}(g^2)$ corrections to $\Zg$.

\section{Thermal modes at $\OO(g^2)$}
\la{sec:O:gsq:calculation}

We embark on the calculation of the $T$-scale contribution to $\Zg$ at NLO.
In principle, this
requires the computation of the two-loop diagrams $(b)$--$(j)$ in Fig.~\ref{fig:diagrams}.
As we are dealing with the temperature scale,
there is no need to or benefit in using any IR-resummed
propagators and vertices 
in the real-time formalism.
The unresummed nature of the propagators can be used to our advantage to
greatly simplify the calculation.

In a nutshell, all but diagrams $(c),(f),(g)$ only 
contain terms that are proportional to either
$\bigl\langle A^- A^-\bigr\rangle$ or
$\bigl\langle A^- A^x\bigr\rangle$, 
both of which vanish in Feynman gauge. 
Furthermore, following an argument to be given in 
Sec.~\ref{subsec:eom}, we will see
that only diagram $(c)$ needs to be computed.
The remainder of the section is then devoted to the splitting up and the computation of
the different contributions:
quark and gauge (gluon and ghost) in the polarisation tensor. 

\subsection{Equations of motion of the light-like Wilson line}
\la{subsec:eom}

Looking at Eq.~\eqref{eq:zg_wline}, we can write
\begin{equation}
\label{eq:deffundlorentz}
     \Zg =  \frac{-1}{\dR \CR} \int_0^\infty  \!{\rm d}x^+ x^+
     \,\Tr\,\Bigl\langle
       U_\rmii{R}(-\infty;x^+) F^{-\perp} (x^+)\,
       U_\rmii{R}(x^+;0)\,F^{-\perp}(0)
       U_\rmii{R}(0;-\infty) \Bigr\rangle
      \,,
\end{equation}
where $F^{-\perp}F^{-\perp}=F^{-x}F^{-x}+F^{-y}F^{-y}$.
Continuing onwards, let us recast Eq.~\eqref{eq:deffundlorentz} as a double integral, using translation invariance
\begin{align}
\label{eq:deffunddouble}
     \Zg &=
      \frac{-1}{\dR\CR}
        \int_0^\infty \! {\rm d} x^+
        \int_0^{x^+} \! {\rm d} y^+
     \nn &
     \hphantom{{}=\frac{-1}{\dR\CR}}
     \times
     \Tr\,\Bigl\langle
       U_\rmii{R}(-\infty;x^++y^+) F^{-\perp} (x^++y^+)\,
       U_\rmii{R}(x^++y^+;y^+)\,
       F^{-\perp}(y^+)
       U_\rmii{R}(y^+;-\infty) \Bigr\rangle
       \,,
       \nn &=
      \frac{-1}{\dR\CR}
      \int_0^\infty \! {\rm d} x^+
      \int_0^{\frac{x^+}{2}} \! {\rm d} y^+
     \,\Tr\,\Bigl\langle
       U_\rmii{R}(-\infty;x^+) F^{-\perp} (x^+)\,
       U_\rmii{R}(x^+;y^+)\,
       F^{-\perp}(y^+)
      U_\rmii{R}(y^+;-\infty)
      \Bigr\rangle
      \,,
\end{align}
where we have shifted $x^+$ in going to the second line.
We now rewrite $\Zg$ as
\begin{equation}
    \Zg\equiv \Zg^{\perp\perp}+\Zg^{\perp-}+\Zg^{-\perp}+\Zg^{--}
    \,,
    \label{zgsplit}
\end{equation}
with the index labelling to be elaborated on momentarily.
By calling on a method introduced in~\cite{Caron-Huot:2008dyw}, we rewrite
\begin{equation}
F^{-\perp}=-\partial^\perp A^-+[D^-,A^\perp]
  \,.
\end{equation}
The equation of motion of the Wilson line,
$D^-_{x^+} U(x^+;y^+)=0$ can then be used to rewrite
the commutator as a total derivative, i.e.
\begin{equation}
\label{totald}
  U_\rmii{R}(a;x^+)\bigl[D^-,A^\perp(x^+)\bigr]U_\rmii{R}(x^+;b)=
  \frac{{\rm d}}{{\rm d}x^+}\bigl[U_\rmii{R}(a;x^+)\,A^\perp(x^+)\,U_\rmii{R}(x^+;b)\bigr]
  \,.
\end{equation}
This should shed light on the labelling in Eq.~\eqref{zgsplit}:
the indices there denote whether the $-\partial^\perp A^-$ (${}^\perp$) or the $[D^-,A^\perp]$
component (${}^-$) of the field strength tensor has been taken.

We start with the first term from Eq.~\eqref{zgsplit},
\begin{equation}
  \Zg^{\perp\perp}=-\frac{1}{\dR\CR} \int_0^\infty \!{\rm d} x^+ x^+
     \,\Tr\,\Bigl\langle
       U_\rmii{R}(-\infty;x^+) \partial^{\perp}A^- (x^+)\,U_\rmii{R}(x^+;0)\,
       \partial^{\perp}A^-(0)U_\rmii{R}(0;-\infty)     \Bigr\rangle.
       \label{perpperp}
\end{equation}
For the second term, we have, employing Eq.~\eqref{totald}
\begin{align}
  \Zg^{\perp-}&=\frac{1}{\dR\CR} \int_0^\infty\!{\rm d} x^+
     \,\Tr\,\Bigl\langle
       U_\rmii{R}(-\infty;x^+)\partial^{\perp}A^- (x^+)\,U_\rmii{R}(x^+;x^+/2)\,
    \nn &\hphantom{{}=\frac{1}{\dR\CR}}
       \times
       \Big[
          A^\perp(x^+/2)U_\rmii{R}(x^+/2;0)
        - U_\rmii{R}(x^+/2;0)A^\perp(0)
       \Big]
      U_\rmii{R}(0;-\infty)
    \Bigr\rangle
    \,.
\end{align}
Translation invariance dictates that for the first part,
\begin{align}
    &\Bigl\langle
      U_\rmii{R}(-\infty;x^+)\partial^{\perp}A^- (x^+)\,
      U_\rmii{R}(x^+;x^+/2)\, A^\perp(x^+/2)
      U_\rmii{R}(x^+/2;-\infty)\Bigr\rangle
      \nn =&
    \Bigl\langle
      U_\rmii{R}(-\infty;x^+/2)\partial^{\perp}A^- (x^+/2)\,
      U_\rmii{R}(x^+/2;0)\,A^\perp(0)
      U_\rmii{R}(0;-\infty)\Bigr\rangle
    \,,
\end{align}
so that, by aptly redefining the integration variable,
one finds, including its counterpart, that
\begin{align}
\label{perpminus}
  \Zg^{\perp-}&=\frac{1}{\dR\CR} \int_0^\infty \!{\rm d} x^+ %
     \,\Tr\,\Bigl\langle
    U_\rmii{R}(-\infty;x^+)\partial^{\perp}A^- (x^+)\,
    U_\rmii{R}(x^+;0)\,A^\perp(0)
    U_\rmii{R}(0;-\infty)
  \Bigr\rangle
  \,, \\[1mm]
   \Zg^{-\perp} &= 
      \frac{-1}{\dR\CR} \int_0^\infty\!{\rm d}x^+ 
     \,\Tr\,\Bigl\langle
      U_\rmii{R}(-\infty;x^+) A^{\perp} (x^+)\,U_\rmii{R}(x^+;0)\,
      \partial^{\perp}A^-(0)      U_\rmii{R}(0;-\infty)     \Bigr\rangle
      \label{minusperp}
  \,,
\end{align}
where we used the fact that, in a non-singular gauge, $A^\perp(\infty)=0$
and we applied translation invariance and relabeled $y^+$ into $x^+$.
Finally, for the term with both commutators, we find
\begin{align}
\label{minusminus}
    \Zg^{--}=\frac{1}{\dR\CR}
     \,\Tr\,\Bigl\langle
       U_\rmii{R}(-\infty;0)A^\perp (0)\,
       A^\perp(0)
      U_\rmii{R}(0;-\infty)
    \Bigr\rangle
    \,,
\end{align}
where we have again set the boundary term at infinity to zero.

At two-loop level, Eq.~\eqref{perpperp} can only
receive a contribution from diagram $(c)$ in Feynman gauge;
terms with an $A^-$ pulled from the Wilson line cannot contribute,
as there is no triple $A^-$ vertex. Moreover,
there cannot be a contribution of the form of diagram $(g)$ as 
the $gA^{-}A^{\perp}$ part of the covariant derivative 
has been removed from the correlator through the previous manipulations.
Similar arguments can be safely applied to Eqs.~\eqref{perpminus} and \eqref{minusperp}:
they only receive a ``mixed'' contribution from diagram $(c)$; this relies on 
the observation that no vertex with two $A^-$ gluons exists.

Eq.~\eqref{minusminus} is the only piece of the decomposition that does not vanish at LO
in Feynman gauge; it corresponds precisely to diagram $(a)$.
Eq.~\eqref{minusminus} also receives an NLO contribution from $(c)$. It also contains the only contribution directly 
involving the Wilson lines, as either of them can source an $A^-$ gluon 
that can connect with a three-gluon vertex to the two $A^\perp$ ones. However,
that would give rise to the following colour structure 
\begin{equation}    
    \mathcal{C}=f^{abc}
      A^{\perp\,a}
      A^{\perp\,b}
      A^{-\,c}
    \,,
\end{equation}
which vanishes, due to the symmetric (anti-symmetric) nature of $A^{\perp\,a}A^{\perp\,b}$ ($f^{abc}$).
Therefore, through this reorganisation, we conclude that all of the NLO corrections to $\Zg$ come from diagram $(c)$.
We arrive at the same conclusion
from an explicit real-time computation of diagrams $(f)$ and $(g)$, whose
contributions end up cancelling each other~\cite{eamonnthesis}.

\subsection{Decomposition of the gluon self-energy}
\la{subsec:decomp}

We thus dive straight into the computation of diagram $(c)$
\begin{equation}
    \Zg^{(c)} = 
    -\frac{(D-2)}{\dR\CR}\int_0^\infty\! {\rm d} x^+\,x^+ 
    \Tr\Bigl\langle
      (\partial^{-}A^{x}(x^+)-\partial^{x}A^{-}(x^{+}))
      (\partial^{-}A^{x}(0)-\partial^{x}A^{-}(0))
    \Bigr\rangle
    \,,
\end{equation}
where we have used rotational invariance in the transverse plane. As the field-strength insertions are at positive
time separation, there is indistinguishably a forward Wightman or time-ordered one-loop 
propagator stretching between them 
\begin{align}
  \Zg ^{(c)}&=
  -(D-2)\int_0^\infty\!{\rm d} x^+\int_{Q}x^+e^{-iq^-x^+}\Big[
        (q^{-})^2G_{>}^{xx}(Q)
      - 2q^{x}q^{-}G_{>}^{x-}(Q)
      + (q^{x})^{2}G_{>}^{--}(Q)\Big]
    \nn &=
    (D-2)\int_{Q}\frac{1+\nB(q^0)}{(q^{-}-i\varepsilon)^2}(F_{\rmii{$R$}}(Q)-F_{\rmii{$A$}}(Q))
    \,,
\end{align}
where we have used the Kubo--Martin--Schwinger (KMS) relation to write
the  forward Wightman~$>$ propagator in terms of the statistical function $1+\nB(q^0)$ and
the one-loop spectral function
$F_{\rmii{$R$}}(Q)-F_{\rmii{$A$}}(Q)$.
Its retarded and advanced components read
\begin{align}
  F_{\rmii{$R$/$A$}}(Q) &=\frac{i}{(Q^2\pm i\varepsilon q^0)^2}\Big[
          (q^{-})^2\Pi^{xx}_{\rmii{$R$/$A$}}(Q)
        - 2q^{x}q^{-}\Big(\Pi^{0x}_{\rmii{$R$/$A$}}(Q)-\Pi^{zx}_{\rmii{$R$/$A$}}(Q)\Big)
      \nn &\hphantom{{}=\frac{i}{(Q^2\pm i\varepsilon q^0)^2}\Big[}
      + (q^x)^{2}\Big(
            \Pi^{00}_{\rmii{$R$/$A$}}(Q)
          + \Pi^{zz}_{\rmii{$R$/$A$}}(Q)
          - 2\Pi^{z0}_{\rmii{$R$/$A$}}(Q)\Big)
      \Big]
    \,,
    \label{eq:f_before_proj}
\end{align}
when written in terms of the one-loop self-energies, given in Appendix~\ref{app:self}.
It is then useful to decompose the self-energies in terms of their
longitudinal ($L$) and
transverse ($T$) components (with respect to the three-momentum), which can also be found in Appendix~\ref{app:self}
\begin{equation}
    \Pi^{\rmii{$R$/$A$}}_{\mu\nu}(Q)=
     P^{\rmii{$L$}}_{\mu\nu}(Q)\Pi^{\rmii{$R$/$A$}}_{\rmii{$L$}}(Q)
    +P^{\rmii{$T$}}_{\mu\nu}(Q)\Pi^{\rmii{$R$/$A$}}_{\rmii{$T$}}(Q)
    \,
    \label{eq:selfenergydecomp}
\end{equation}
where
$P^{\rmii{$L$}}_{\mu\nu}$ and
$P^{\rmii{$T$}}_{\mu\nu}$ are given in Eqs.~\eqref{eq:lprojector}--\eqref{eq:tprojector}.

We can then rewrite Eq.~\eqref{eq:f_before_proj} as
\begin{equation}
    F_{\rmii{$R$/$A$}}(Q)=
    \frac{i}{(Q^2\pm i\varepsilon q^0)^2}\Big[
          (q^{-})^2\Pi_{\rmii{$T$}}^{\rmii{$R$/$A$}}(Q)
        + \frac{(q^x)^{2}Q^2}{q^2}\Big(
            \Pi_{\rmii{$L$}}^{\rmii{$R$/$A$}}(Q)
          - \Pi_{\rmii{$T$}}^{\rmii{$R$/$A$}}(Q)
    \Big)\Big]
  \,,
\end{equation}
which leaves us with the decomposition
\begin{align}
    \Zg^{(c)}&=
      \int_{Q}\frac{1+\nB(q^0)}{(q^- - i\varepsilon)^2}\bigg\{
    \frac{i}{(Q^2+i\varepsilon q^0)^2}\Big[
          (D-2)(q^{-})^2\Pi_{\rmii{$T$}}^{\rmii{$R$}}(Q)
        + \frac{q_{\perp}^{2}Q^2}{q^2}\Big(
            \Pi_{\rmii{$L$}}^{\rmii{$R$}}(Q)
          - \Pi_{\rmii{$T$}}^{\rmii{$R$}}(Q)
        \Big)\Big]
    -\text{adv.}\bigg\}
  \nn[2mm] &=
    Z_{\mathrm{g},q^{-2}}^{(c)}
  + Z_{\mathrm{g},\rmii{$L$-$T$}}^{(c)}
  \,,\
  \label{eq:split}
\end{align}
where adv.\ stands for the advanced part while
$Z_{\mathrm{g},q^{-2}}^{(c)}$
corresponds to the $(q^{-})^2$-
and
$Z_{\mathrm{g},\rmii{$L$-$T$}}^{(c)}$ to the
$\Pi_\rmii{$L$}-\Pi_\rmii{$T$}$-proportional
terms
\begin{align}
    Z_{\mathrm{g},q^{-2}}^{(c)}&=\int_{Q}[1+\nB(q^0)]\bigg\{
        \frac{i(D-2)\Pi_{\rmii{$T$}}^{\rmii{$R$}}(Q)}{(Q^2+i\varepsilon q^0)^2}
      - \text{adv.}\bigg\}
    \,,\label{defqm}
    \\
    Z_{\mathrm{g},\rmii{$L$-$T$}}^{(c)}&= \int_{Q}\frac{1+\nB(q^0)}{(q^- - i\varepsilon)^2}\bigg\{
      \frac{i}{(Q^2+i\varepsilon q^0)}\frac{q_{\perp}^{2}}{q^2}\Big(\Pi_{\rmii{$L$}}^{\rmii{$R$}}(Q)-\Pi_{\rmii{$T$}}^{\rmii{$R$}}(Q)\Big)
    - \text{adv.}\bigg\}
  \,.\label{deflt}
\end{align}
The strategy adopted for the remainder of this section is to compute $\rmO(g^2)$ corrections to $\Zg$ according to the above decomposition. Rather than wading through technical details (which can be found in Appendix~\ref{app:cdetails}), we instead focus on highlighting and discussing the physical relevance of the various divergences that we encounter.

\subsection{$ Z_{\mathrm{g},q^{-2}}^{(c)}$ contribution}
\la{subsec:qms}
For the computation of $ Z_{\mathrm{g},q^{-2}}^{(c)}$,
we can resort to standard Euclidean techniques.
We start from
\begin{equation}
  Z^{(c)}_{\mathrm{g},q^{-2}}= i(D-2)\int_{Q}\Bigl(\frac12+\nB(q^0)\Bigr)\bigg[
      \frac{\Pi_\rmii{$T$}^{\rmii{$R$}}(Q)}{(Q^2+i\varepsilon q^0)^2}
    - \frac{\Pi_\rmii{$T$}^{\rmii{$A$}}(Q)}{(Q^2-i\varepsilon q^0)^2}
    \bigg],
    \label{zcexprstarteucl}
\end{equation}
where we have kept the even-in-$q^0$ part of the integrand only.
By analyticity, we have
\begin{equation}
    Z^{(c)}_{\mathrm{g},q^{-2}}=-(D-2)\Tint{\QE}
    \frac{\Pi_\rmii{$T$}^{\rmii{E}}(\QE)}{\QE^4}
    \,.
    \label{zcexpreucl}
\end{equation}
Here, $\QE^2\equiv q_0^2+q^2$ is the Euclidean four-momentum squared,
with $q_0=2\pi n T$ a bosonic Matsubara frequency and $\Tinti{\QE}$
the standard sum-integral, see Eq.~\eqref{def_integrals}. Likewise,
$\Pi_\rmii{$T$}^{\rmii{E}}(\QE)$ is the transverse Euclidean polarisation tensor, given
in Appendix~\ref{app:self}.
The integrations can then be carried out using standard Euclidean techniques,
as detailed in Appendix~\ref{app:eucl}.

The contribution of a gluon and ghost loop reads (cf.\ Eq.~\eqref{zcexpreuclmasterbosonnonzero})
\begin{align}
    Z^{(c)}_{\mathrm{g},q^{-2},\rmii{B},n\ne0}=&
    \frac{g^2 \CA T^2}{96\pi^2}\bigg[
    \frac{11}{3}\bigg(\frac{1}{\epsilon}
    +4 \ln\frac{\bmu}{T}\bigg)
    \nn &
    +8 \ln A
    -4 \gammaE
    +\frac{61}{9}
    +\frac{10}{3}\ln\pi
    -\frac{52}{3} \ln 2
    +\frac{4}{3}\times\bigl[0.299378\bigr]
    \bigg]
  \,,\label{zcqmsqmbosonnonzero}
\end{align}
where B stands for bosonic, $\gammaE$ is the Euler-Mascheroni constant and
$A \approx 1.28243$ is the Glaisher constant.
Here and in all following formulae we truncate the expansion in $D=4-2\epsilon$
to the order-$\epsilon^0$ term included. 
We have further subtracted off the $q_0=0$ zero-mode contribution associated with the $\QE$ sum-integral. 
This IR-divergent zero-mode contribution can be found in Sec.~\ref{sec:matching:to:EQCD} 
(see Eq.~\eqref{IRdivqm}), where we cancel its divergence with 
an opposite one from the EQCD subtraction term discussed in and after Eq.~\eqref{eq_subtr}.

The fermionic contribution from the quark loop reads --- see Eq.~\eqref{zcexpreuclmasterfermion}
\begin{align}
    Z^{(c)}_{\mathrm{g},q^{-2},\rmii{F}}=-\frac{g^2 \Tf\Nf T^2}{96\pi^2}\bigg[
    \frac43\bigg(\frac{1}{\epsilon}
      + 4 \ln\frac{\bmu A^3}{4\pi T}\bigg)
      + 4 \gammaE
      + \frac{26}{9}
      - \frac{10}{3}\times \bigl[ 0.199478 \bigr]
    \bigg].\label{zcqmsqmfermion}
\end{align}
In this case,
the $\QE$ zero mode is included, as its contribution is finite as expected from
the purely non-Abelian nature of the $c_2$ subtraction in
Eqs.~\eqref{eq_subtr}--\eqref{subtrterms}.

Upon summing Eqs.~\eqref{zcqmsqmbosonnonzero} and \eqref{zcqmsqmfermion},
we find
\begin{align}
  Z^{(c),\rmi{IR safe}}_{\mathrm{g},q^{-2}} &=
        Z^{(c)}_{\mathrm{g},q^{-2},\rmii{B},n\ne0}
      + Z^{(c)}_{\mathrm{g},q^{-2},\rmii{F}}
    \nn[2mm] &=
    \frac{g^2 T^2}{96\pi^2}\bigg[
        \Big(\frac{11}{3}\CA-\frac43\Tf\Nf\Big)
        \Big(\frac{1}{\epsilon}+4 \ln\frac{\bmu}{T}\Big)
    \nn &\hphantom{{}=\frac{g^2 T^2}{96\pi^2}\bigg[}
    + \CA\Big(
        8\ln A
      - 4 \gammaE
      + \frac{61}{9}
      + \frac{10}{3}\ln\pi
      - \frac{52}{3}\ln 2
      + \frac{4}{3}\times \bigl[0.299378\bigr]\Big)
    \nn &\hphantom{{}=\frac{g^2 T^2}{96\pi^2}\bigg[}
    - \Tf\Nf\Big(
        16\ln A 
      + 4\gammaE
      + \frac{26}{9}
      - \frac{16}{3}\ln(4\pi)
      - \frac{10}{3}\times\bigl[0.199478\bigr]\Big)
    \bigg]
  \,.\label{zcqmsqm}
\end{align}

The attentive reader will have noticed that Eqs.~\eqref{zcqmsqmbosonnonzero} and \eqref{zcqmsqmfermion} are divergent.
The very attentive reader will also have noticed that the divergent term in Eq.~\eqref{zcqmsqm} is proportional to
the first coefficient of the QCD $\beta$ function.
Indeed, the divergence is of ultraviolet origin:
it arises from the thermal part of a bosonic loop integral multiplying the vacuum part of the other loop integral.
This divergence is then removed by
charge renormalisation. In more detail, let us consider the standard relation between bare 
and renormalised coupling at one-loop level,
{\em viz.}\ 
\begin{equation}
    \label{counterterm}
    g^2_\mathrm{bare}=
    g^2(\bmu)\bigg[1-\frac{g^2b_0}{(4\pi)^2\epsilon} \bigg]
    + \mathcal{O}(g^6)
    \,,\quad\text{where}\quad
    b_0=\frac{11}{3}\CA-\frac43\Tf\Nf
    \,.
\end{equation}
Let us then go back to the leading-order term in Eq.~\eqref{zglo}, recalling that, as per Eq.~\eqref{eff_mass},
$\minf^2$ is obtained by multiplying $\Zg$ by $g^2$.%
\footnote{%
\label{foot_gsquare}We are following the notation originally introduced in~\cite{CaronHuot:2008uw}; arguably,
moving $g^2$ into the definition of $\Zg$ %
would be preferable, as it would make %
this operator
scheme-independent. 
}
Hence, if that $g^2$ is replaced by Eq.~\eqref{counterterm}, 
the $\mathcal{O}(g^4)$ term there generates the counterterm
\begin{align}
    \Zg^{\rmii{LO},\rmi{ct}}=&-(d-1)\frac{g^2b_0}{(4\pi)^2\epsilon}
    \int_\vec{q}\frac{\nB(q)}{q}
    =\frac{g^2  T^2}{96\pi^2}
    \Big(\frac{11}{3}\CA-\frac43\Tf\Nf\Big)
    \Big[-\frac{1}{\epsilon}-2 \ln\frac{\bmu A^{12}}{4\pi T}+1\Big]
    \,,
    \label{countertermeval}
\end{align}
when multiplying the LO term in Eq.~\eqref{zglo}.
Upon adding Eqs.~\eqref{zcqmsqm} and~\eqref{countertermeval},
we find%
\footnote{%
  The fermionic piece
of the expression below has also been determined using real-time methods such as those employed in the
next subsection, see~\cite{eamonnthesis}.
The two determinations agree within numerical uncertainty.
}
\begin{align}
    Z^{(c),\rmi{IR/UV}\,\rmi{safe}}_{\mathrm{g},q^{-2}} &=
    Z^{(c),\rmi{IR}\,\rmi{safe}}_{\mathrm{g},q^{-2}}
    +\Zg^{\rmii{LO},\rmi{ct}}
    \nn[2mm] &=
    \frac{g^2 T^2}{96 \pi^2}\bigg\{
   \Big(\frac{11}{3}\CA-\frac43\Tf\Nf\Big)
   2\ln\frac{\bmu_\rmii{UV}}{T}
   \nn &
   +\CA \bigg[
    - 80 \ln A
    - 4 \gammaE
    + \frac{94}{9}
    + \frac{32}{3} \ln\pi
    - \frac{8}{3} \ln 2
    + \frac{4}{3}\times\bigl[0.299378\bigr]\bigg]
  \nn &
  + \Tf\Nf \bigg[
      16\ln A
    - 4\gammaE
    - \frac{38}{9}
    + \frac{8}{3}\ln (4\pi)
    + \frac{10}{3}\times \bigl[0.199478\bigr] \bigg]\bigg\}
  \,.\label{zcqmsqmfinal}
\end{align}

Here,
we relabeled $\bmu \to \bmu_\rmii{UV}$ to distinguish it from other occurrences of $\bmu$, which 
are to be understood as factorisation scales, whereas $\bmu_\rmii{UV}$ is a genuine renormalisation
scale. 
When merging the perturbative and lattice data, we will fix $g^2$ to the values
that correspond to those chosen in the various lattice ensembles, as explained in Appendix~\ref{app:coupling}.

\subsection{$Z_{\mathrm{g},\rmii{$L$-$T$}}^{(c)}$ contribution}
\label{sec:LT}

In this subsection, differently from the previous one, we rely on real-time methods:
as detailed in Appendix~\ref{sec:gluon_lmt}, standard Euclidean 
techniques are not applicable for this class of integrals, 
due to the  $(q^--i\varepsilon)^2$ denominator in Eq.~\eqref{deflt}.
We thus set up everything so that the integration over $K$, the loop momenta running through the self-energy blob is done in
four dimensions. 
We choose this scheme to avoid having to consider the $D$-dimensional equivalents 
of the one-loop self-energies listed in Appendix~\ref{app:self}. 
The integration over $Q$, the momentum picked up during the forward scattering,
is instead done in $D=4-2\epsilon$ dimensions.As we shall
see, this contribution will present divergences showing sensitivity to
hard, collinear and soft regions of phase space. When dealing with these 
regions one then needs to adopt the same scheme, as we shall do in Sec.~\ref{sec:matching:to:EQCD}
for the $L-T$ part of the soft contribution.

The fermionic and bosonic $L-T$ contributions are computed in Appendix~\ref{app:fermion_lmt} and~\ref{sec:gluon_lmt}, respectively. 
Furthermore, as the vacuum contribution
of the longitudinal and transverse self-energies is identical by Lorentz invariance, no charge renormalisation 
is needed in this case; indeed, the counterterm~\eqref{countertermeval} was fully absorbed 
by the $(q^{-})^2$ contribution.

Our result then comes from the sum of Eqs.~\eqref{finalLTfermion_app} and~\eqref{finalLTgluon_app},
{\em viz.}
\begin{align}
  Z_{\mathrm{g},\rmii{$L$-$T$},n\neq 0} ^{(c)}&=\frac{\mD^2}{(4\pi)^2}
    \bigg\{\frac{2}{\epsilon^2}
    + \frac{2}{\epsilon} \Bigl(
      \ln\frac{\bmu e^{\gammaE-\frac{1}{2}}}{2\pi T}%
    + \ln\frac{\bmu A^{12}e^{-\frac{1}{2}}}{8\pi T}
    \Bigr)
    \nn &\hphantom{{}=\frac{\mD^2}{(4\pi)^2}\bigg\{}
    + 2 \ln^2\frac{\bmu e^{\gammaE-\frac12}}{2\pi T}-4 \gamma_1
    +\frac{\pi^2}{4}-2 \gammaE^2+\frac32-2(\ln2-1)^2
    \nn &\hphantom{{}=\frac{\mD^2}{(4\pi)^2}\bigg\{}
    + 2 \ln^2\frac{\bmu A^{12}e^{-\frac12}}{8 \pi T}
    + \frac{1}{2}\ln^2(2)
    - 2(\ln\zeta_2)'^{2}
    + \frac{2\zeta_2''}{\zeta_2}
    \nn &\hphantom{{}=\frac{\mD^2}{(4\pi)^2}\bigg\{}
    + \frac12\mbox{Li}_2\Bigl(\frac{2}{3}\Bigr)
    + \frac14\mbox{Li}_2\Bigl(\frac{3}{4}\Bigr)
    - \frac12
    - \frac14 \ln(3) \ln\frac{4}{3}
    \bigg\}
    \nn &
    + \frac{g^2 \CA T^2}{48\pi^2}\bigg\{0.145277-5.13832\bigg\}
    \nn &%
    -\frac{g^2 \Tf \Nf T^2}{48\pi^2} \bigg\{
        \frac{2\ln2 }{\epsilon}
      + 4\ln(2)\ln\frac{\bmu A^{12}e^{-\frac12}}{8\pi T}
      + 2\ln^2(2)
      + 0.438129
      - 1.97382
   \bigg\}\,,
    \label{finalLTtot}
\end{align}
where $\zeta_s = \zeta(s)$ is the Riemann zeta function,
$(\ln \zeta_s)' = \zeta'(s)/\zeta(s)$ and
$\gamma_1$ is the first Stieltjes constant.
For the $\CA$-proportional gauge contribution,
we have again subtracted off the zero-mode associated with the $Q$ integration.
Its value is given in Eq.~\eqref{IRdivLT}.
The parts that have been obtained numerically have not been summed, to highlight their 
origin from the various contributions listed in Appendix~\ref{app:fermion_lmt} and~\ref{sec:gluon_lmt}.
What remains contains a $1/\epsilon^2$ pole, which signals the presence of double logarithmic divergences
not of vacuum origin.
These divergences originate from a \emph{pole term} and a \emph{cut term}. The former denotes 
the contribution to Eq.~\eqref{deflt} where the discontinuity implied in the ``retarded minus advanced''
spectral structure comes from the pole in the $1/Q^2$ propagator. The latter instead
acquires spectral weight from the imaginary parts of the polarisation tensor. 

To clarify the physical origin of these divergences, we now inspect these divergent terms,
deferring their precise evaluation to Appendix~\ref{app:fermion_lmt} and~\ref{sec:gluon_lmt}.
For what concerns the pole term, it arises from Eq.~\eqref{zcexprLTHTLpole} and
its bosonic equivalent,
yielding
\begin{equation}
  Z_{\mathrm{g},\rmii{$L$-$T$},\rmi{pole}}^{(c)}=
     - \frac{\mD^2}{2}\int_{Q}\frac{\frac12+\nB(|q^0|)}{(q^--i\varepsilon)^2}\frac{q_\perp^2}{q^2 }2\pi\delta(Q^2)
    =- \frac{\mD^2}{2}\int_{\vec{q}}\frac{1+2\nB(q)}{2q^3}\frac{q_\perp^2+2q_z^2}{q_\perp^2 }
    \,.
    \label{zcexprLTHTLpole2div}
\end{equation}
While the vacuum term would be scale-free and vanishing in DR, it physically corresponds
to a UV divergence for $q^z\to \infty$, which is accompanied by a \emph{collinear divergence}.
This shows up for $q^z\gg q_\perp\to 0$ , i.e.\footnote{%
There is in addition a power-law divergent zero-mode term from the $T/q^0$ infrared
limit of the Bose--Einstein distribution. This is not unexpected: it 
corresponds to the $\mathcal{O}(\mD^2)$ term in the $q\gg \mD$
expansion of Eq.~\eqref{loeqcd}. It too needs subtracting, as it is already included
there. DR takes care of this automatically.}
\begin{equation}
  Z_{\mathrm{g},\rmii{$L$-$T$},\rmi{pole div}}^{(c)}=
  -\mD^2
    \int_{\vec{q}}\frac{1}{2\vert q^z\vert}\frac{1}{q_\perp^2 }\bigg\vert_{q^z\gg q_\perp}
    \,.
    \label{zcexprLTHTLpole2div2}
\end{equation}
While we leave the proper analysis of this divergence to future work, we think it is safe 
to speculate that it will be absorbed by properly dealing with the small $q_\perp$ and large $q^z$
regions. The former will require systematic resummation of the asymptotic mass 
and of multiple soft scattering effects, likely relating this double-logarithmic divergence to 
the ones observed in the transverse-momentum broadening coefficient in \cite{Liou:2013qya,Blaizot:2013vha}
and systematised for a weakly-coupled QGP in \cite{Ghiglieri:2022gyv}.\footnote{It also remains 
to be understood how such a long-duration collinear process could be used as an ingredient 
in the determination of an equally long-duration (formation time) medium-induced radiation. This question 
too is left for future work.} The large-$q^z$ region is 
instead likely related to the emergence of the $q^z\sim E$ region, with $E\gg T$ the large 
light-cone energy of the hard jet parton, thus relating this divergence to the 
possible loop corrections to the matching coefficients in Eq.~\eqref{eff_mass}.

Let us now inspect the cut term.
As detailed in Appendix~\ref{app:fermion_lmt} and~\ref{sec:gluon_lmt},
the divergence arises again from the vacuum part, i.e.\ $\theta(q^0)$, in the $1+\nB(q^0)$ prefactor.
As explained there, it is convenient to make the integrand 
explicitly even in $q^0$,
leading to
\begin{align}
    \label{zcexprlogssymvacdivtot}
    Z_{\mathrm{g},\rmii{$L$-$T$},\rmi{cut div}}^{(c)}&=
    \frac{g^2}{16\pi}
    \int_0^\infty\!{\rm d}k \, f(k) \int_0^\infty\frac{{\rm d}q^0}{2\pi}
    \int_{\vec{q}}\frac{q_\perp^2}{q^5}\bigg[\frac{1}{(q^0-q^z-i\varepsilon)^2}
    +\frac{1}{(q^0+q^z+i\varepsilon)^2}\bigg]
    \nn
    &\times
  \theta(q-2k)\bigg[
      \left(3(2k-q^0)^2-q^2\right)\theta(q+2k-q^0)\theta(q^0-q)
    \nn & \hphantom{\times\theta(q-2k)}
    - \left(3(2k+q^0)^2-q^2\right)\theta(q^0-q+2k)\theta(q-q^0)
    \bigg]
    \,,
\end{align}
where 
\begin{equation}
    \label{deffk}
    f(k)\equiv 2\Tf \Nf \nF(k)+\CA \nB(k)
    \,,
\end{equation}
and $\nF(k)\equiv(\exp(k/T)+1)^{-1}$ is the Fermi--Dirac distribution.
Eq.~\eqref{zcexprlogssymvacdivtot} arises from appropriately combining Eqs.~\eqref{zcexprLTHTLcutsympos}
and \eqref{zcexprlogssymvac}, together with their bosonic counterparts like Eq.~\eqref{zcexprlogssymvacsympg}.
We only consider here the $\theta(q-2k)$ part  of Eqs.~\eqref{zcexprlogssymvac} and \eqref{zcexprlogssymvacsympg}
as it is the divergent one.

Let us consider for illustration the simplest part of the above,
the $q^2$-proportional piece
in $3(2k- q^0)^2-q^2)$ in the time-like slice $q^0>q$.
The main message will not be altered by including the other terms.  
It reads
\begin{align}
    \label{illustration}
    \int_0^\infty\!{\rm d}k \,f(k)  &
    \int_{\vec{q}}\int_{q}^{q+2k}\frac{{\rm d}q^0}{2\pi}
    \frac{q_\perp^2}{q^3}\bigg[\frac{1}{(q^0-q^z)^2}
    +\frac{1}{(q^0+q^z)^2}\bigg] \theta(q-2k)
    \nn
    &=2\int_0^\infty\!{\rm d}k \,k\, f(k) \int_{\vec{q}}\frac{2(k q+q_z^2)+q_\perp^2}{\pi q^3(q_\perp^2
    +4 k q +4k^2)} \theta(q-2k)
    \\
    \label{illint}
    &=
    \int_0^\infty\!{\rm d}k \,k\, f(k) \int_{\vec{q}_\perp}\frac{4 q_\perp^2}{\pi^2 (q_\perp^2+4k^2)^2}\left[\ln\frac{q_\perp}{2k}+\mathcal{O}(1)+\mathcal{O}(k/q_\perp)\right]
    \,,
\end{align}
where the arising logarithmic term is what causes the double pole in DR.
Its $k$-dependent argument is what is responsible for the different arguments of
the logarithms in the fermionic and bosonic case, 
see the $\Nf$-proportional $\ln2$ term in Eq.~\eqref{finalLTtot}.
Moreover, the $\ln(q_\perp/(2k))$ must come from 
two regions of the $q^z$ integral in Eq.~\eqref{illustration}, once we take the asymptotic UV 
expansion $q_\perp\gg k$.
Indeed, we have
a first region for
$q^z\sim q_\perp \gg k$ and
a second region for
$q_z\sim q_\perp^2/k\gg q_\perp$ --- we are exploiting evenness in $q^z$ to consider
twice the positive range only. We then have, by appropriately expanding 
the integrand in the two regions and separating them by a cutoff $\mu$
\begin{align}
    \int_0^\mu \frac{{\rm d}q^z}{\pi}\frac{2q_z^2+q_\perp^2}{\pi q^3q_\perp^2}&=
    \frac{\ln\frac{4\mu^2}{q_\perp^2}-1}{\pi^2q_\perp^2}
    \,,
    &{\rm and}&&
    \int_\mu^\infty \frac{{\rm d}q^z}{\pi} 
    \frac{2}{\pi q^z(q_\perp^2+4 k q^z )} &=
    \frac{\ln\frac{q_\perp^4}{16 k^2\mu^2}}{\pi^2q_\perp^2}
    \,,
\end{align}
which reproduces the $q_\perp\gg k$ limit of Eq.~\eqref{illint}.

This then clarifies that the cut term has a double-logarithmic UV divergence, which we can again 
speculate to be related to the emergence of the $q^+\sim E$ region. Its proper handling is also left
to future work. We conclude by remarking that in the real-time calculation of 
$ Z_{\mathrm{g},q^{-2}}^{(c)}$ (cf.~\cite{eamonnthesis}) there is a cancellation 
between a double-log akin to that in Eq.~\eqref{illint} 
and an opposite one arising from a pole contribution which is very different 
from the simple one in Eq.~\eqref{zcexprLTHTLpole2div2}. The collinear sensitivity of the latter
and its absence in $ Z_{\mathrm{g},q^{-2}}^{(c)}$ --- see~\cite{eamonnthesis} --- might explain why no such cancellation happens for
$Z_{\mathrm{g},\rmii{$L$-$T$}}^{(c)}$.

\section{Cancellation of IR QCD divergences with UV EQCD divergences}
\la{sec:matching:to:EQCD}

\begin{figure}[t]
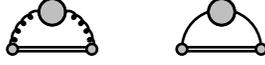

\centering
  \begin{align*}
    \ToptWSBB(\Lsri,\Agliii,\Agliii,\Asai,\Asai)\;
    \qquad
    \ToptWSBB(\Lsri,\Asai,\Asai,\Asai,\Asai)\;
  \end{align*}
\caption{%
    EQCD equivalent of diagram $(c)$. The labelling is identical
    to that from Fig.~\ref{fig:diagrams} except that here, curly lines
    represent spatial gauge bosons,
     double lines are adjoint Wilson lines and
    solid lines are adjoint scalars.
    }
\label{fig:c}
\end{figure}

As we explained around Eq.~\eqref{eq_subtr}, the EQCD evaluation in \cite{Ghiglieri:2021bom} relied 
on a subtraction of the $c_2$-proportional UV-log divergent term from NLO perturbative EQCD 
in Eq.~\eqref{eq_subtr}, which introduced 
a logarithmic dependence on the $L_\mathrm{min}$ regulator. The $Q$ zero modes in the perturbative 
QCD evaluation of the previous section, on the other hand, introduced an 
IR log-divergence,
{\em viz.}
\begin{align}
    Z^{(c)}_{\mathrm{g},q^{-2},\rmii{B},n=0}&=
    \frac{3g^2 \Nc T^2}{64\pi^2}\bigg[
    -\frac{1}{\epsilon}-4\ln \left(\frac{\bmu}{2 T}\right)-\frac{5}{3}
    \bigg],
    \label{IRdivqm}
    \\
    Z^{(c)}_{\mathrm{g},\rmii{$L$-$T$},\rmii{B},n=0}&=
    \frac{g^2\CA T^2}{32\pi^2}\left[-\frac{1}{\epsilon}-2\ln\left(\frac{\bmu }{4T}\right)+2\right]
    \,,
    \label{IRdivLT}
\end{align}
which have been obtained from Eqs.~\eqref{zcexpreuclmasterbosonzeromode} and \eqref{LmTzeromodediv}.

Our goal in this section is to 
add back a DR version of the NLO subtraction in Eq.~\eqref{eq_subtr}. 
The resulting UV poles in $1/\epsilon$ 
should cancel against those in Eqs.~\eqref{IRdivqm} and \eqref{IRdivLT}. 
To this end, we note that \cite{Ghiglieri:2021bom} used Feynman gauge  for the 3D gauge field of EQCD.
From the arguments of Sec.~\ref{subsec:eom} 
we then know that this $c_2$-proportional UV divergence must come from
the EQCD analogue of diagram~$(c)$, shown in Fig.~\ref{fig:c}.
This suggests that
\begin{equation}
     \Zg^\text{match} \equiv
      Z^{(c)}_{\mathrm{g},q^{-2},\rmii{B},n=0}
     + Z^{\rmi{3d}\,(c)}_{\mathrm{g},q^{-2},c_2}
     + Z^{(c)}_{\mathrm{g},\rmii{$L$-$T$},\rmii{B},n=0}
     + Z^{\rmi{3d}\,(c)}_{\mathrm{g},\rmii{$L$-$T$},c_2}
     \,,
     \label{defzgmatch}
\end{equation}
where
$Z^{\rmi{3d}\,(c)}_{\mathrm{g},q^{-2},c_2}$ is the $(q^{-})^{2}$ and
$Z^{\rmi{3d}\,(c)}_{\mathrm{g},\rmii{$L$-$T$},c_2}$ the $L-T$ piece
of the UV divergence, evaluated in DR.
When adding Eq.~\eqref{defzgmatch} to~\eqref{eq_subtr},
we will undo the $c_2$ subtraction and thus arrive at
\begin{equation}
    \Zg^\text{non-pert class}=
      Z_{\mathrm{g}}^{\rmi{3d}}\big\vert^\text{merge}
    + \Zg^\text{match}
    \,,
    \label{deftotmatch}
\end{equation}
which is a finite,
$L_\mathrm{min}$-independent result for the non-perturbative evaluation of the classical contribution.
The best-available determination of $Z_g$ is then 
\begin{align}
  \label{deftotmatchsumLO}
  \Zg^{\text{non-pert class+NLO }T}&=
  \Zg^\rmii{LO}
  +\delta\Zg^{\text{non-pert class+NLO }T}
  \,,\\[2mm]
  \delta\Zg^{\text{non-pert class+NLO }T}&=
      \Zg^\text{non-pert class}
    + Z^{(c),\rmi{IR/UV safe}}_{\mathrm{g},q^{-2}}
    + Z_{\mathrm{g},\rmii{$L$-$T$},n\neq 0}^{(c)}
    \,,
    \label{deftotmatchsum}
\end{align}
where 
$\delta\Zg^{\text{non-pert class+NLO }T}$
incorporates the classical modes to all 
orders and the thermal modes to NLO. The final two terms in Eq.~\eqref{deftotmatchsum} have been given in Eqs.~\eqref{zcqmsqmfinal} and \eqref{finalLTtot}. 

Let us now begin the determination of
$Z^{\rmi{3d}\,(c)}_{\mathrm{g},q^{-2},c_2}$ and
$Z^{\rmi{3d}\,(c)}_{\mathrm{g}\,\rmii{$L$-$T$},c_2}$, starting from the former.
With respect to the decomposition defined in Eq.~\eqref{eq:zg3d},
we first take the $\langle BB\rangle$ contribution from
Appendix A.3 of~\cite{Ghiglieri:2021bom} (cf.\ Eq.~(A.14))
\begin{equation}
   \langle BB\rangle_{\text{div}}^{(c)}=\frac{1}{4}\int_{\vec{q}}\frac{e^{iLq_z}}{q^4}(q_{\perp}^2+(d-1)q_{z}^2)\delta^{ab}\Big(-\Pi^{\rmii{$T$}}_{A^a_iA^b_j\,\text{div}}(q)\Big)
   \,,
    \label{eq:eqcddivT}
\end{equation}
where $\Pi^{\rmii{$T$}}_{A^a_iA^b_j\,\text{div}}$ is the EQCD transverse gluon polarisation tensor in the massless limit.
As we are investigating a UV divergence on the EQCD side, the $q\gg \mD$ limit is the appropriate one.
It can be extracted either from the $\mD\to 0$ limit of Eq.~(A.10) in~\cite{Ghiglieri:2021bom}
or from the zero-mode contribution to Eq.~\eqref{mikkopol}. We find
\begin{eqnarray}
\label{eq:bb_selfDR}
  \Pi^{\rmii{$T$}}_{A^a_iA^b_j\,\rmi{div}}(q) &=&
    \delta^{ab}
    \frac{3\gE^2 \CA}{(4\pi)^\frac{d}{2}}
    \frac{
      \Gamma \bigl( 1 - \frac{d}{2} \bigr)
      \Gamma \bigl(\frac{d}{2}\bigr)^2}{
      \Gamma \bigl( d-1 \bigr)}
      \mu^{3-d}
      q^{d-2}
    \,,\\
\label{eq:bb_self}
    &\stackrel{d=3-2\epsilon}{=}&
    -\delta^{ab}\frac{3\gE^2\CA q}{16}
    \,,
\end{eqnarray}
where we have denoted the $d=3$ limit for future use.

The second term in Eq.~\eqref{eq:eqcddivT},
proportional to $(d-1)q_z^2$, 
corresponds to the
$Z^{\rmi{3d}\,(c)}_{\mathrm{g},q^{-2},c_2}$ contribution.
To obtain it, we must then integrate in $L$ up to the cutoff while performing 
the $q$ integration in DR, so that its $\Zg^\text{match}$ contribution 
undoes the corresponding piece of the $c_2$ subtraction. 
Looking at Eq.~\eqref{eq:zg3d} and using Eq.~\eqref{eq:bb_selfDR},
we then find 
\begin{align}
  Z^{\rmi{3d}\,(c)}_{\mathrm{g},q^{-2},c_2} &=
  - \frac{4T}{\dA}\int_{0}^{L_{\text{min}}}\! {\rm d}L\,L\,\langle BB\rangle_{q^{-2}\,\text{div}}
  =\frac{3\gE^2\CA T }{128\pi^{d}}
  \frac{
  \bigl(d-1\bigr)
  \bigl(2 d-5\bigr)}{
  \bigl( d-3 \bigr)
  }
  \frac{
  \Gamma \bigl(\frac{d}{2}-2\bigr)
  \Gamma \bigl(\frac{d}{2}-1\bigr)}{
  \bigl(L_{\rmi{min}}\mu \bigr)^{2d-6}
  }
   \nn &=
   \frac{3\gE^2\CA T}{64\pi^2}\bigg[
        \frac{1}{\epsilon}
      - 7
      + 4\gammaE
      + 4\ln({L_{\text{min}}\bar{\mu}})%
   \bigg]
   \,.
    \label{EQCDcqmsquare}
\end{align}
Since $\gE^2= g^2T+\mathcal{O}(g^4)$, the $1/\epsilon$ term cancels against the one from Eq.~\eqref{IRdivqm}.

As for the $L-T$ piece, we need
the $q_\perp^2$-proportional piece of Eq.~\eqref{eq:eqcddivT}
together with the contribution from the $\langle EE \rangle$ correlator. We 
extract this from Eq.~(A.8) of~\cite{Ghiglieri:2021bom} and expand towards the UV to arrive at 
\begin{eqnarray}
\label{PiLEQCD}
    \Pi_{\Phi^a\Phi^b}(q) &=&
      \delta^{ab}\frac{\gE^2
      \CA\mD}{16\pi}\left(-4-\frac{6\CF-\CA}{6\CA}\frac{\lambdaE}{\gE^2}-8\frac{q^2-\mD^2}{q \mD}\arctan\frac{q}{\mD}\right)
    \nn &\stackrel{q\gg \mD}{\approx}&
    -\delta^{ab}\frac{\gE^2\CA q}{4}\equiv\Pi_{\Phi^a\Phi^b\,\text{div}}(q)
    \,.
\end{eqnarray}
As the corresponding IR pole has been obtained using the mixed scheme described at the beginning of Sec.~\ref{sec:LT},
we stick to the same scheme here and have evaluated Eq.~\eqref{PiLEQCD} in $d=3$.
From Eq.~(A.13) of \cite{Ghiglieri:2021bom}, we then write
\begin{equation}
    \langle EE\rangle_{\text{div}}^{(c)}=\frac{1}{4}\int_{\vec{q}}\frac{e^{iL q_z}}{q^4}q_{\perp}^2\delta^{ab}
    \Big(-\Pi^{\rmii{$T$}}_{\Phi^a\Phi^b\,\text{div}}(q)\Big)
    \,.
    \label{eq:eqcddivL}
\end{equation}
To get the full $\Pi_{\rmii{$L$}}-\Pi_{\rmii{$T$}}$ EQCD contribution,
we need to add to this its $\langle BB\rangle$
counterpart.
It arises from the $q_\perp^2$-proportional piece in Eq.~\eqref{eq:eqcddivT} with 
the $d=3$ polarisation tensor in Eq.~\eqref{eq:bb_self}. We then 
evaluate the Fourier transform in DR and carry out the $L$ integration, as in Eq.~\eqref{EQCDcqmsquare}.
This yields --- see again Eq.~\eqref{eq:zg3d}
\begin{align}
  Z^{\rmi{3d}\,(c)}_{\mathrm{g}\,\rmii{$L$-$T$}\,c_2}&=
  \frac{4T}{\dA}\int_{0}^{L_{\text{min}}}\!{\rm d}L\,L\,\Bigl(
      \langle EE\rangle_{\rmi{div}}^{(c)}
    - \langle BB\rangle_{q_\perp^2\,\rmi{div}}^{(c)}
    \Bigr) =
  \frac{\gE^2 \CA T}{(4\pi)^{\frac{d+1}{2}}}
  \frac{
    \Gamma\bigl(\frac{d+1}{2}\bigr)}{
    \bigl(3-d\bigr)}
  \biggl( \frac{L_\rmi{min}\mu}{2} \biggr)^{3-d}
  \nn &=
  \frac{\gE^2\CA T}{32\pi^2}\left[
      \frac{1}{\epsilon}
    - 1+2\gammaE
    + 2\ln\frac{{L_{\text{min}}\bmu}}{2}%
  \right]
  \,,
\label{EQCDcLTdiv}
\end{align}
which, reassuringly contains a $1/\epsilon$ pole that cancels exactly against the one from Eq.~\eqref{IRdivLT}.
We note that, if the Fourier transforms in Eqs.~\eqref{eq:eqcddivT} and \eqref{eq:eqcddivL} are carried 
out in three dimensions, the former vanishes, whereas the latter gives
$\langle EE\rangle_{\text{div}\,d=3}^{(c)}=\CA\dA\gE^2/(4\pi L)^2$, from which the value of $c_2$ in Eq.~\eqref{subtrterms}
is extracted.
We have split our calculation in
a $(q^{-})^{2}$ part and
a $L-T$ part, rather than
a $\langle EE\rangle$ and
a $\langle BB\rangle$ one,
to check the explicit cancellation of $1/\epsilon$ poles in each of the former 
two parts and to account for our mixed scheme in the $L-T$ contribution.

We can finally insert Eqs.~\eqref{IRdivqm}, \eqref{IRdivLT}, \eqref{EQCDcqmsquare} and \eqref{EQCDcLTdiv} into
Eq.~\eqref{defzgmatch}, together with the LO EQCD matching $\gE^2=g^2T$, to find
\begin{equation}
    \Zg^\text{match}=\frac{ g^2 \CA T^2}{8 \pi ^2}\bigg[2 \ln (2L_{\text{min}} T)+2 \gammaE -3\bigg]
    \,.
    \label{sumup}
\end{equation}
Note that in computing $\Zg^\text{non-pert class}$, the $\ln L_{\text{min}}$ above cancels against the
$-\int \frac{{\rm d}L}{L} c_2 \gE^2$ term in Eq.~\eqref{eq_subtr}, as we originally set to show.

We conclude by noting that Eq.~\eqref{sumup} is somewhat arbitrary: 
a firm necessity of the cancellation of the $L_\text{min}$ dependence of
$Z_{\mathrm{g}}^{\rmi{3d}}\big\vert^\text{merge}$
through the IR divergence of the thermal scale contribution is the $\ln(L_\text{min} T)$ term. The non-logarithmic
parts, on the other hand, can freely be shuffled between Eq.~\eqref{sumup} and Eqs.~\eqref{zcqmsqmfinal} and \eqref{finalLTtot}
by redefining the scheme used to separate the IR-divergent part from the remainder in Sec.~\ref{sec:O:gsq:calculation}.
This is of course irrelevant once these three equations are summed in Eq.~\eqref{deftotmatchsum}.
However, in the next section we shall not include the $Z_{\mathrm{g}\,\rmii{$L$-$T$}\,n\neq 0} ^{(c)}$
contribution, as it contains extra divergences signaling sensitivities to new regions of phase space
we have left to future work. Hence, to present a first estimate of the impact of neglecting 
 $Z_{\mathrm{g}\,\rmii{$L$-$T$}\,n\neq 0} ^{(c)}$, let us
determine what would happen to Eq.~\eqref{sumup} if we were to omit the finite parts
associated with the $L-T$ contribution. By dropping them from Eqs.~\eqref{IRdivLT} and \eqref{EQCDcLTdiv}
we find 
\begin{equation}
\label{sumupfiniteparts}
    Z_{\mathrm{g},\rmi{no $L$-$T$ finite parts}}^\text{match}=
      Z_{\mathrm{g}}^\text{match}
    - \frac{ g^2 \CA T^2}{32 \pi ^2}\bigg[1+2\ln2+2\gammaE\bigg]
    \,.
\end{equation}
The numerical impact of Eqs.~\eqref{sumup} and~\eqref{sumupfiniteparts} will be discussed in the next section.

\section{Summary and discussion}
\la{sec:discussion}
In Section~\ref{sec:O:gsq:calculation}, we determined the contribution to $\Zg$
of the temperature scale at order $g^2$: it is given by  Eqs.~\eqref{zcqmsqmfinal}
and \eqref{finalLTtot} for the two contributions in which it has 
naturally been split; see Eq.~\eqref{eq:split}. 
In Sec.~\ref{sec:matching:to:EQCD}, we analyzed the IR divergences of the $T$-scale contribution
and showed how they cancel against the UV divergences from EQCD found in 
\cite{Ghiglieri:2021bom} when they are consistently evaluated in the same DR scheme. 
In Eq.~\eqref{deftotmatch}, we explain how to add the lattice result of~\cite{Ghiglieri:2021bom}
with the finite remainder, Eq.~\eqref{sumup}, of the cancellation of the divergences, lifting
the artificial dependence on the UV cutoff introduced in~\cite{Ghiglieri:2021bom}.

To evaluate the numerical impact of our work,
we should
in principle use our best available result, summing the NLO $T$-scale contribution 
with the all-order non-perturbative evaluation.
This is given by Eq.~\eqref{deftotmatchsum},
which, after inserting all ingredients,
gives rise to
\begin{align}
  \delta \Zg^{\text{non-pert class+NLO }T}&=
    Z_{\mathrm{g}}^{\rmi{3d}}\big\vert^\text{merge}
  + \frac{g^2\CA T^2}{8 \pi ^2}\bigg[2 \ln (2L_{\text{min}} T)+2 \gammaE -3\bigg]
  \nn &
    + \frac{g^2 T^2}{96\pi^2}\bigg\{
      \bigg(\frac{11}{3}\CA-\frac43\Tf\Nf\bigg)2\ln\frac{\bmu_\rmii{UV}}{T}
      \nn &
   + \CA \bigg[
      - 80\ln A
      - 4 \gammaE
      + \frac{94}{9}
      + \frac{32}{3}\ln\pi
      - \frac{8}{3}\ln 2
      + \frac{4}{3}\times\bigl[0.299378\bigr]\bigg]
    \nn &
    + \Tf\Nf \bigg[
        16\ln A
      - 4\gammaE
      - \frac{38}{9}
      + \frac{8}{3}\ln (4\pi)
      + \frac{10}{3}\times\bigl[0.199478\bigr]
    \bigg]\bigg\}
  \nn &
  + \frac{\mD^2}{(4\pi)^2}\bigg\{
      \frac{2}{\epsilon^2}
    + \frac{2}{\epsilon} \Bigl(
        \ln\frac{\bmu e^{\gammaE-\frac{1}{2}}}{2\pi T}%
      + \ln\frac{\bmu A^{12}e^{-\frac{1}{2}}}{8\pi T}
      \Bigr)
  \nn &\hphantom{{}+ \frac{\mD^2}{(4\pi)^2}\bigg\{}
    + 2 \ln^2\frac{\bmu e^{\gammaE-\frac{1}{2}}}{2\pi T}
    - 4 \gamma _1
    + \frac{\pi ^2}{4}
    - 2 \gammaE^2
    - 1 +4\ln2
  \nn &\hphantom{{}+ \frac{\mD^2}{(4\pi)^2}\bigg\{}
    + 2 \ln^2\frac{\bmu A^{12}e^{-\frac{1}{2}}}{8\pi T}
    - \frac{3}{2}\ln^2(2)
    - 2(\ln\zeta_2)'^{2}
    + \frac{2\zeta_2''}{\zeta_2}
  \nn &\hphantom{{}+ \frac{\mD^2}{(4\pi)^2}\bigg\{}
    + \frac12\mbox{Li}_2\Bigl(\frac{2}{3}\Bigr)
    + \frac14\mbox{Li}_2\Bigl(\frac{3}{4}\Bigr)
    - \frac14 \ln(3) \ln\frac{4}{3}
    \bigg\}
    \nn &
    + \frac{g^2 T^2}{48\pi^2}\bigg\{
      \CA\bigg[0.145277-5.13832\bigg]
    \nn &%
    - \Tf\Nf\bigg[
      \frac{2\ln2}{\epsilon}
    + 4\ln(2)\ln\frac{\bmu A^{12}e^{-\frac12}}{8\pi T}-2\ln^2(2)
    + 0.438129
    - 1.97382
  \bigg]\bigg\}
  \,.
  \label{finaltot}
\end{align}

The one outstanding roadblock 
is that our evaluation of the ``longitudinal minus transverse''
$Z_{\mathrm{g},\rmii{$L$-$T$}\,n\neq 0}^{(c)}$ contribution in
Eq.~\eqref{finalLTtot} contains yet uncancelled double-logarithmic divergences 
from the sensitivity to the
collinear ($q_\perp\sim \minf$) and
hard ($q^+\sim E\gg T$) 
regions of phase space. We have left their understanding and cancellation to future work.
Here, we present a numerical analysis of the impact of the other corrections, 
namely
the $(q^{-})^2$ contribution,
$Z^{(c),\rmi{IR/UV safe}}_{\mathrm{g},q^{-2}}$,
given in Eq.~\eqref{zcqmsqmfinal} and
the  classical contribution determined from lattice and
perturbative EQCD in~\cite{Ghiglieri:2021bom}
complemented by the cancellation of the logarithmic sensitivity achieved through
Eqs.~\eqref{defzgmatch} and \eqref{sumup}.
More precisely,
let us define 
\begin{align}
\label{deffinite}
    \delta \Zg^\text{finite}&\equiv
      \Zg^\text{non-pert class}
    + Z^{(c),\rmi{IR/UV safe}}_{\mathrm{g},q^{-2}}
    \\&=
    Z_{\mathrm{g}}^{\rmi{3d}}\big\vert^\text{merge}
    + \frac{g^2 \CA T^2}{8\pi^2}\bigg[2 \ln\frac{T}{2\gE^2}+2 \gammaE -3\bigg]
    \nn &
    + \frac{g^2 T^2}{96 \pi^2}\bigg\{
     \Big(\frac{11}{3}\CA-\frac43\Tf\Nf\Big)2\ln\Bigl(4\pi e^{-\gammaE}\exp\frac{-\Nc + 4\ln(4)\Nf}{22 \Nc - 4\Nf}\Bigr)
  \nn &
  +\CA \bigg[
    - 80\ln A
    - 4 \gammaE
    + \frac{94}{9}
    + \frac{32}{3} \ln\pi
    - \frac{8}{3}\ln 2
    + \frac{4}{3}\times\bigl[0.299378\bigr]
  \bigg]
  \nn &
  + \Tf\Nf \bigg[
      16\ln A
    - 4\gammaE
    - \frac{38}{9}
    + \frac{8}{3}\ln (4\pi)
    + \frac{10}{3}\times\bigl[0.199478\bigr]
  \bigg]\bigg\}
  \,,\label{totfinite}
\end{align}
which is obtained by summing Eqs.~\eqref{zcqmsqmfinal} and \eqref{defzgmatch}. We also set
$\bmu_\rmii{UV}$ to Eq.~\eqref{fixmuuv}, in keeping with the renormalisation scale used in the EQCD matching,
and $L_\text{min}=1/(4\gE^2)$, the value used in $Z_{\mathrm{g}}^{\rmi{3d}}\big\vert^\text{merge}$ in \cite{Ghiglieri:2021bom}.
To further motivate 
our choice of looking, for the time being, at this partial subset of $\OO(g^2)$ corrections, 
let us remark 
that $Z^{(c)\text{ IR/UV safe}}_{\mathrm{g},q^{-2}}$ comes from the same $(q^{-})^2$ 
structure as the LO term --- compare Eq.~\eqref{zglo} and \eqref{defqm}. 
It is thus fundamentally different from
$Z_{\mathrm{g},\rmii{$L$-$T$},n\neq 0} ^{(c)}$, 
which indeed contains a new structure and new divergences.

\begin{table}[t]
    \centering
    \begin{tabular}{|r|l|l|l||l|}
    \hline
        \multicolumn{1}{|c|}{$T$} &
        \multicolumn{1}{|c|}{$\displaystyle\frac{\Zg^\text{non-pert class}}{T^2}$} &
        \multicolumn{1}{|c|}{$\displaystyle\frac{\delta\Zg^\text{finite}}{T^2}$} &
        \multicolumn{1}{|c||}{$\displaystyle\frac{\delta Z_{\mathrm{g},\rmii{Eq.\eqref{sumupfiniteparts}}}^{\rmi{finite}\vphantom{X^2}}}{T^2}$} &
        \multicolumn{1}{c|}{$\displaystyle\frac{Z_{\mathrm{g},\rmii{LO}}^{\rmi{3d}}}{T^2}$}
        \\[8pt]\hline
        $250$~MeV   %
        &$-0.513(138)(45)(7)$
        &$-0.340(138)(45)(7)$ 
        &$-0.465(138)(45)(7)$
        &$-0.376$\\
        $500$~MeV %
        &$-0.619(99)(39)(3)$
        &$-0.491(99)(39)(3)$
        &$-0.584(99)(39)(3)$
        &$-0.324$\\
        $1$~GeV%
        &$-0.462(71)(9)(7)$
        &$-0.356(71)(9)(7)$
        &$-0.430(71)(9)(7)$
        &$-0.305$\\
        $100$~GeV %
        &$-0.327(16)(5)(2)$
        &$-0.274(16)(5)(2)$
        &$-0.310(16)(5)(2)$
        &$-0.223$\\
        \hline
    \end{tabular}
    \caption{Values for our main results.
    $\delta Z_{\mathrm{g},\rmii{Eq.\eqref{sumupfiniteparts}}}^\rmi{finite}$
    is defined by replacing Eq.~\eqref{sumup} with Eq.~\eqref{sumupfiniteparts} in $\Zg^\text{non-pert class}$ and
    $Z_{\mathrm{g},\rmii{LO}}^{3\text{d}}$ is given by Eq.~\eqref{loeqcd}.
    See Appendix~\ref{app:coupling} for details on the renormalisation scale and the values of the coupling.
    We recall that these values are to be added to the LO value $\Zg^\rmii{LO}=T^2/6$, as per Eq.~\eqref{deftotmatchsumLO}.}
    \label{tab:full_values}
\end{table}
In Table~\ref{tab:full_values},
we collect the numerical values for our main results.
We refer to Appendix~\ref{app:coupling}
for the coupling prescriptions
and the EQCD data for $Z_{\mathrm{g}}^{\rmi{3d}}\big\vert^\text{merge}$ from \cite{Ghiglieri:2021bom}. 
$\delta Z_{\mathrm{g}\,\rmii{Eq.\eqref{sumupfiniteparts}}}^\text{finite}$
is defined by replacing Eq.~\eqref{sumup} with Eq.~\eqref{sumupfiniteparts} in
$\Zg^\text{non-pert class}$, 
thus giving a first, very rough estimate of the error associated with the lack of inclusion of the $L-T$
contribution. The quoted uncertainties are inherited from \cite{Ghiglieri:2021bom}. They come from statistical errors in lattice EQCD (first bracket), from the integration of the IR tail in Eq.~\eqref{defzgmatch} (second bracket) and from the quadrature of the 
lattice data (third bracket).

As we see, our results for $\delta \Zg^\text{finite}$ are negative and, in magnitude, larger than $\Zg^\rmii{LO}=T^2/6$ (see Eq.~\eqref{zglo}),
to which they are to be added, as per Eq.~\eqref{deftotmatchsumLO}.
With the exception of the highest temperature, 
within errors they are however compatible with the $\OO(g)$ correction only, i.e.\
the leading-order contribution in perturbative EQCD given by
$Z_{\mathrm{g},\rmii{LO}}^{\rmi{3d}}$ in the last column of the table.
Recall that
$Z_{\mathrm{g},\rmii{LO}}^{\rmi{3d}}$ is obtained in Eq.~\eqref{loeqcd}, which in turn clarifies why
a negative contribution arises:
it comes from removing the incorrect, unscreened IR of bare QCD
and replacing it with the correct, screened IR of EQCD. As the screened contribution is necessarily
smaller than the bare unscreened one, negativity sets in. 

This large negative contribution, resulting in an overall negative $\Zg$, may appear problematic. 
However, until the divergences of the $L-T$ contributions are addressed by properly including
collinear and hard  modes, we refrain from making definite statements 
on the convergence of EQCD factorisation for this observable. Already by simply reshuffling 
the finite parts associated with the cancellation of IR and UV divergences in the $L-T$ channel ---
the $\delta Z_{\mathrm{g},\rmii{Eq.\eqref{sumupfiniteparts}}}^\text{finite}$ column --- we see a significant
variation. Fully understanding these extra regions of phase space may give rise to even larger
contributions, due to potentially large double logarithms.

These potentially enhanced double logarithms would relate
the $T$-scale physics in Eqs.~\eqref{finaltot}
and Eq.~\eqref{finalLTtot} with hard, $q^+\sim E\gg T$ physics related to the
EFT matching implicit in Eq.~\eqref{eff_mass}, as we commented there. They would also 
include collinear physics, with effects such as the quantum-mechanical interference 
of multiple soft scatterings, as we mentioned in Sec.~\ref{sec:LT} after Eq.~\eqref{zcexprLTHTLpole2div2}.
At the moment we do not speculate further on these matters and leave them to future work.

As a final remark, we comment on the expansion undergirding the 
factorisation of the asymptotic mass into the operators $\Zg$ and $Z_\mathrm{f}$
described by Eq.~\eqref{eff_mass}. It requires the parton energy $E$
to be $E\gg T$ when 
the $T$-scale contribution to these operators
is considered. If the soft contribution is considered,
the expansion becomes $E\gg \mD$, so that, in the weak-coupling 
regime, it remains applicable for $E\sim T$. This also
holds for Eqs.~\eqref{deftotmatch} and \eqref{sumup},
as the latter is obtained from the soft limit
of the $T$-scale contribution.\footnote{%
We remark that, even in this regime, our results remain different 
from those of~\cite{Gorda:2022fci,Ekstedt:2023anj,Ekstedt:2023oqb,Gorda:2023zwy}. These papers determine the 
two-loop and power-suppressed one-loop contributions from the hard thermal scale to the self-energy for soft external modes, to 
then take the asymptotic mass limit. It then corresponds to first expanding for $P\ll Q_i\sim T$, where $P$ is the external 
momentum and $Q_i$ the loop momenta, to then take the $P\gg \mD$ limit. Our calculation, on the other hand, corresponds to
taking a $P\gg Q_i\sim T$ expansion to then take the $P\to T\gg \mD$ limit from above.
Indeed, our calculation in this limit sees IR
divergences given by Eqs.~\eqref{IRdivqm}--\eqref{IRdivLT} which are then absorbed by EQCD, whereas the results of~\cite{Gorda:2022fci,Ekstedt:2023anj,Ekstedt:2023oqb,Gorda:2023zwy} are finite after charge renormalization, further highlighting their difference.} This has then the very 
important consequence that our values for $\Zg^\text{non-pert class}$,
tabulated in Tab.~\ref{tab:full_values}, are valid
also when considering partons with thermal energies. This 
then makes them applicable for corrections to 
observables such as the thermal photon rate or transport
coefficients, where they could be used to complement the
existing NLO determinations \cite{Ghiglieri:2013gia,Ghiglieri:2018dib,Ghiglieri:2018dgf}.

\section*{Acknowledgements}
We thank Guy Moore for useful conversations.
JG, PS, NS and EW acknowledge support by a PULSAR grant from the R\'egion Pays de la Loire. JG acknowledges support 
by the Agence Nationale de la Recherche under grant ANR-22-CE31-0018 (AUTOTHERM). 
PS acknowledges support by the Deutsche Forschungsgemeinschaft (DFG, German Research Foundation) through
the CRC-TR 211 `Strong-interaction matter under extreme conditions' --
project number 315477589 --
TRR 211.

\appendix
\renewcommand{\thesection}{\Alph{section}}
\renewcommand{\thesubsection}{\Alph{section}.\arabic{subsection}}
\renewcommand{\theequation}{\Alph{section}.\arabic{equation}}

\section{Conventions}\label{sec_conv}
Our sign for the covariant derivative is
\begin{equation}
    D_{\mu}=\partial_{\mu}-igA_{\mu}
    \,,\nonumber
\end{equation}
which fixes the sign of the three gluon vertex to be positive.
Moreover, we employ the ``mostly minus" $(+,-,-,-)$ metric.
Uppercase letters denote four-momenta, lowercase letters the modulus
of the three-momenta.
When using light-cone coordinates
\begin{align}
  p^{+} &\equiv\frac{p^{0}+p^{z}}{2}=\bar{v}\cdot P
  \,,&
  p^{-} &\equiv p^{0}-p^{z}=v\cdot P
  \,,&
  P\cdot Q &=p^{+}q^{-}+p^{-}q^{+}-\bpp\cdot \bqp
  \,,
\end{align}
where the two light-like reference vectors are defined as
\begin{equation}
    \bar{v}^{\mu}\equiv\frac{1}{2}(1,0,0,-1)
    \,,\qquad
    v^{\mu}\equiv(1,0,0,1)
    \,.\nonumber
\end{equation}
This asymmetric convention for the ${}^+$ and ${}^-$ components of the light-cone 
coordinates has two advantages: it has unitary Jacobian, i.e.\
${\rm d}p^0 {\rm d}p^z={\rm d}p^+{\rm d}p^-$,
in scalings where $p^-\ll p^+$ it then implies
$p^0\approx p^z\approx p^+$.

The emergent momenta integrals
are often regularised in $d=D-1=3-2\epsilon$,
employing the $\overline{\mbox{MS}}$ renormalisation scale
$\bmu^{2} = 4\pi e^{-\gammaE}\mu^{2}$.
Denoting spatial momenta as
$\vec{p}=(\vec{p}_{\perp}^{ },p_{z}^{ })$,
this allows to use the shorthand
\begin{align}
  \label{def_integrals}
  \Tint{P_\rmii{E}}&\equiv T\sum_{n\in \mathbb{Z}}\int_{\vec p}\Big\vert_{p_0=2n \pi T}
  \,,&
  \int_{P}&\equiv
  \mu^{3-d}\int\frac{{\rm d}^D P}{(2\pi)^D}
  \,,&
  \int_{\vec{p}}&\equiv
  \mu^{3-d}\int\frac{{\rm d}^d \vec{p}}{(2\pi)^d}
  \,,
  \nn
  \Tint{\{P_\rmii{E}\}} &\equiv T\sum_{n\in \mathbb{Z}}\int_{\vec p}\Big\vert_{p_0=(2 n-1)\pi T}
  \,,&
  \int_{\vec{p}_{\perp}}&\equiv
  \mu^{3-d}\int\frac{{\rm d}^{d-1} \vec{p}_{\perp}^{ }}{(2\pi)^{d-1}}
  \,.
\end{align}

\section{One-loop self-energies and projectors}\label{app:self}

We document in this appendix the self energies that are needed for
the computation of diagram $(c)$ in Sec.~\ref{sec:O:gsq:calculation}. 
We start by providing the transverse and longitudinal projectors (with respect to the three-momentum) that are used
to arrive at the decomposition Eq.~\eqref{eq:split}
\begin{align}
  \label{eq:lprojector}
  P_\rmii{$L$}^{\mu\nu}(Q)&=
    - g^{\mu\nu}
    + \frac{q^{\mu}q^{\nu}}{Q^2}
    - P_\rmii{$T$}^{\mu\nu}(Q)
  \,,\\
    \label{eq:tprojector}
  P_\rmii{$T$}^{\mu\nu}(Q)&=
    - g^{\mu\nu}
    + u^{\mu}u^{\nu}
    - \frac{
        q^{\mu}q^{\nu}
      - (Q\cdot u)q^{\mu}u^{\nu}
      - (Q\cdot u)q^{\nu}u^{\mu}
      + (Q\cdot u)^2 u^{\mu}u^{\nu}}{q^2}
  \,,\\[2mm]
  \label{eq:bathframe}
  (u^{\mu})&=(1,0,0,0)
  \,.
\end{align}

The main ingredient for the Euclidean evaluation of the $(q^{-})^2$ contribution 
is the 
Feynman-gauge transverse Euclidean polarisation tensor, which can
be obtained from, e.g.~\cite{Laine:2016hma}.
It reads
\begin{align}
\Pi_\rmii{$T$}^{\rmii{E}}(Q)&=
    g^2 \Nc\Tint{\KE} \frac{
    (D-2)\KE^2
    -\QE^2
    -\frac{1}{2}\frac{\QE^2}{q^2}(\QE^2+2k^2-2k_0^2)}{\KE^2(\QE-\KE)^2}
    \nn &
    -2g^2 \Tf \Nf \Tint{\{\KE\}} \frac{
    2\KE^2
    -\frac{D-4}{D-2}\QE^2
    -\frac{\QE^2}{(D-2)q^2}(\QE^2+2k^2-2k_0^2)}{\KE^2(\QE-\KE)^2}
    \,.
    \label{mikkopol}
\end{align}

For the $L-T$ contribution,
we need instead the difference between the retarded longitudinal and transverse part,
which appears in Eq.~\eqref{deflt}.
Its gauge contribution (gluon and ghost) and
its quark contribution can be read off from \cite{Kapusta:2006pm},
{\em viz.}
\begin{align}
  \Pi_\rmii{$L$}^{\rmii{$R$}}(Q)_\rmii{B}-\Pi_\rmii{$T$}^{\rmii{$R$}}(Q)_\rmii{B} &=
  \frac{g^2 \CA}{(4\pi)^2 q^3}
   \int_0^\infty\!{\rm d}k \, \nB(k)  \bigg[-8 k q(3 q_0^2-q^2)
   \nn &
   + Q^2\left (12 k^2-q^2+3q_0^2\right) \bigg[
        \ln\frac{q^0+q-2k+i\varepsilon}{ q^0-q-2k+i\varepsilon}
     +  \ln\frac{q^0-q+2k+i\varepsilon}{ q^0+q+2k+i\varepsilon}\bigg]
   \nn &\hspace{-2cm}
   +12 k q^0Q^2 \bigg[
        2\ln \frac{q^0+q+i\varepsilon }{q^0-q+i\varepsilon}
      + \ln\frac{q^0-q+2k+i\varepsilon}{ q^0+q+2k+i\varepsilon}
      - \ln\frac{q^0+q-2k+i\varepsilon}{ q^0-q-2k+i\varepsilon}\bigg]
   \bigg]\,,
   \label{retdiffgluonsimple}
   \\[2mm]
   \Pi_\rmii{$L$}^{\rmii{$R$}}(Q)_\rmii{F}-\Pi_\rmii{$T$}^{\rmii{$R$}}(Q)_\rmii{F} &=  \frac{2 g^2 \Tf\Nf}{(4\pi)^2 q^3}
   \int_0^\infty\!{\rm d}k \, \nF(k)  \bigg[-8 k q(3 q_0^2-q^2)
   \nn &
   + Q^2\left (12 k^2-q^2+3q_0^2\right) \bigg[
        \ln\frac{q^0+q-2k+i\varepsilon}{ q^0-q-2k+i\varepsilon}
      + \ln\frac{q^0-q+2k+i\varepsilon}{ q^0+q+2k+i\varepsilon}\bigg]
   \nn &\hspace{-2cm}
   +12 k q^0Q^2 \bigg[
       2\ln \frac{q^0+q+i\varepsilon }{q^0-q+i\varepsilon}
      + \ln\frac{q^0-q+2k+i\varepsilon}{ q^0+q+2k+i\varepsilon}
      - \ln\frac{q^0+q-2k+i\varepsilon}{ q^0-q-2k+i\varepsilon}\bigg]
   \bigg]
   \,,
   \label{retdiffquarksimple}
\end{align}
where B and F stand for bosonic and fermionic.

\section{Diagram $(c)$ computation: technical details}\label{app:cdetails}

\subsection{Euclidean evaluation of $Z_{\mathrm{g},q^{-2}}$}
\label{app:eucl}
We start from Eq.~\eqref{zcexpreucl}.
Upon inserting Eq.~\eqref{mikkopol} into Eq.~\eqref{zcexpreucl},
we encounter the following class of sum-integrals
\begin{align}
  \mathcal{I}_{s_0;s_1 s_2 s_3}^{\alpha_1 \alpha_2} &=\Tint{\QE \KE}
    \frac{1}{[q^2]^{s_0}}
    \frac{q_0^{\alpha_1} k_0^{\alpha_2}}{
        [\QE^2]^{s_1}
        [\KE^2]^{s_2}
        [(\QE-\KE)^2]^{s_3}}
      \,,\\
  \widehat{\mathcal{I}}_{s_0;s_1 s_2 s_3}^{\alpha_1 \alpha_2} &=\Tint{\QE \{\KE\}}
    \frac{1}{[q^2]^{s_0}}
    \frac{q_0^{\alpha_1} k_0^{\alpha_2}}{
        [\QE^2]^{s_1}
        [\KE^2]^{s_2}
        [(\QE-\KE)^2]^{s_3}}
      \,,
\end{align}
where
$\mathcal{I}_{0;s_1 s_2 s_3}^{00} = \mathcal{I}_{s_1 s_2 s_3}^{ }$ and
fermionic momenta are denoted by curly brackets.
The integrals that appear in Eq.~\eqref{mikkopol} are
\begin{align}
  J_1&=
  \mathcal{I}_{2 0 1}^{ }
      = \Tint{\QE\KE}\frac{1}{\QE^4\KE^2}
      = \mathcal{I}_{1} \mathcal{I}_{2}
    \,,\\[1mm]
   J_2&=
   \mathcal{I}_{1 1 1}^{ }
   =0
   \,,\\[1mm]
    \label{eq:J3}
   J_3&=
   \mathcal{I}_{1;0 1 1}^{ }
    = \Tint{\QE\KE}\frac{1}{\QE^2\KE^2(\vec{q}-\vec{k})^2}
    =\frac{T^2}{192\pi^2}\bigg[\frac{1}{\epsilon}-4\ln\frac{32\pi T}{\bmu A^{12}}+6\bigg]
   \,,\\
   J_4&=
   \mathcal{I}_{1;1 0 1}^{ }
    =
    \Tint{\QE}\frac{1}{q_0^2}\biggl[\frac{1}{q^2} - \frac{1}{\QE^2}\biggr]
    \Tint{\KE}\frac{1}{\KE^2}=
    \frac{2}{(d-2)}\mathcal{I}_{1} \mathcal{I}_{2}
    \,,\\
    \label{eq:J5}
   J_5&=
    \mathcal{I}_{1;1 1 1}^{0 1}
   =\frac{T^2}{288\pi^2}\bigg[\frac{1}{\epsilon}
    +4 \ln\frac{\bmu A^{3}}{4\pi T}
    + 3 \gammaE
    + \frac{95}{12}
    - 0.299378
   \bigg]
   \,,
\end{align}
and the fermionic counterparts $\hat{J}_i$ obtained by summing over odd $\{\KE\}$.
We used the one-loop
bosonic and fermionic master integrals
\begin{align}
  \mathcal{I}_{s} &= \Tint{\PE} \frac{1}{[\PE^2]^s} =
  \mu^{3-d} 2 T \frac{[2\pi T]^{d-2s}}{(4\pi)^\frac{d}{2}}
  \frac{
    \Gamma\bigl( s - \frac{d}{2} \bigr)}{
    \Gamma\bigl( s \bigr)}
  \zeta_{2s -d}
  \,,\\
  \label{eq:I:fer}
  \widehat{\mathcal{I}}_{s} &= \Tint{\{\PE\}} \frac{1}{[\PE^2]^s} =
  (2^{2s-d}-1)\mathcal{I}_{s}
  \,,
\end{align}
where $\zeta_s = \zeta(s)$ is the Riemann zeta function.
$J_2$ vanishes identically, as shown with
integration-by-parts (IBP) methods~\cite{Nishimura:2012ee}.
The integrals $J_3$ in eq.~\eqref{eq:J3} and
$J_5$ in eq.~\eqref{eq:J5} need to be evaluated independently
due to absence of collinearity in the integral~\cite{Davydychev:2022dcw}.
Details on their evaluation are provided below in Appendix~\ref{sub:eucl_integrals}.

The fermionic counterparts read
\begin{align}
  \hat J_1&=
    \widehat{\mathcal{I}}_{2 0 1}^{ }
    = \Tint{\QE\{\KE\}}\frac{1}{\QE^4\KE^2}
    = \widehat{\mathcal{I}}_{1} \mathcal{I}_{2}
   \,,\\[1mm]
   \hat J_2&=
    \widehat{\mathcal{I}}_{1 1 1}^{ }
    = 0
   \,,\\[1mm]
    \label{eq:J3F}
   \hat J_3&=
    \widehat{\mathcal{I}}_{1;0 1 1}^{ }
    = \Tint{\{\QE\KE\}}\frac{1}{\QE^2\KE^2(\vec{q}-\vec{k})^2}
    = \frac{-T^2}{96\pi^2}\bigg[\frac{1}{\epsilon}-4\ln\frac{8\pi T}{\bmu A^{12}}+\ln 4
    \bigg]
   \,,\\
   \hat J_4&=
    \widehat{\mathcal{I}}_{1;1 0 1}^{ }
    = \Tint{\QE}\frac{1}{q_0^2}\biggl[\frac{1}{q^2} - \frac{1}{\QE^2}\biggr]
    \Tint{\{\KE\}}\frac{1}{\KE^2}=
    \frac{2}{(d-2)}\widehat{\mathcal{I}}_{1} \mathcal{I}_{2}
    \,,\\
    \label{eq:J5F}
  \hat J_5&=
  \widehat{\mathcal{I}}_{1;1 1 1}^{0 1}
  =-\frac{5T^2}{1152\pi^2}\bigg[\frac{1}{\epsilon}
    +4 \ln\frac{\bmu}{\pi T}+\frac65 \gammaE +\frac{8}{3}+\frac{168}{5}\ln A+\frac{46}{5}\ln2-0.199478
   \bigg]
   \,.
\end{align}
We used the one-loop fermionic master integrals~\eqref{eq:I:fer} and
that $\hat{J}_2$ too vanishes identically, see e.g.~\cite{Arnold:1994eb}.
Details on $\hat{J}_3$ and $\hat{J}_5$ are provided below in Sec.~\ref{sub:eucl_integrals}.

From Eq.~\eqref{zcexpreucl}, Eq.~\eqref{mikkopol} and the definitions of the master integrals,
we have
\begin{align}
\label{zcexpreuclmaster}
  Z^{(c)}_{\mathrm{g},q^{-2}} =
  -(D-2)g^2\bigg[&
    \Nc\bigg((D-2)J_1-J_2-\frac{J_3}{2}-J_4+2J_5\bigg)
  \nn &
  -2\Tf\Nf \bigg(
        2\hat{J}_1
      - \frac{D-4}{D-2}\hat{J}_2
      - \frac{\hat{J}_3}{D-2}
      - \frac{2\hat{J}_4}{D-2}
      + \frac{4\hat{J}_5}{D-2}
    \bigg)\bigg]
    \,.
\end{align}
The bosonic part gives
\begin{align}
\label{zcexpreuclmasterboson}
    Z^{(c)}_{\mathrm{g},q^{-2},\rmii{B}}&=
    \frac{5g^2\Nc T^2}{576\pi^2}\bigg[
    -\frac{1}{\epsilon}
      - 4 \ln\frac{\bmu}{4\pi T}
      + \frac{48}{5} \ln A
      - \frac{8}{5}\times\bigl[-0.299378\bigr]
      \nn &\hphantom{{}=\frac{5g^2\Nc T^2}{576\pi^2}\bigg[}
      - \frac{24}{5} \gammaE
      - \frac{13}{15}
      - \frac{36}{5}\ln 2
    \bigg]
  \,,\\[2mm]
\label{zcexpreuclmasterbosonzeromode}
   Z^{(c)}_{\mathrm{g},q^{-2},\rmii{B},n=0}&=
   \frac{3g^2\Nc T^2}{64\pi^2}\bigg[
      - \frac{1}{\epsilon}
      - 4\ln \left(\frac{\bmu}{2 T}\right)-
      \frac{5}{3}
    \bigg]
  \,,
\end{align}
where the last line is
the $\QE$ zero-mode part of the $J_i$ integrals, before any $\QE$ shift has been performed.
It gives rise to the soft IR divergence and
can be evaluated relatively straightforwardly (cf.\ e.g. Eq.~\eqref{zeromodemasterboson} and \cite{Brambilla:2010xn}).

After subtracting this zero-mode from the bosonic contribution ---
which we shall match with EQCD later ---
we find
the bosonic and fermionic parts of Eq.~\eqref{zcexpreuclmaster}
\begin{align}
\label{zcexpreuclmasterbosonnonzero}
  Z^{(c)}_{\mathrm{g},q^{-2},\rmii{B},n\ne0} &=
  Z^{(c)}_{\mathrm{g},q^{-2},\rmii{B}}-
  Z^{(c)}_{\mathrm{g},q^{-2},\rmii{B},n=0}
    \nn[2mm]
    &=
    \frac{g^2 \Nc T^2}{96\pi^2}\frac{11}{3}\bigg[
        \frac{1}{\epsilon}
      + 4 \ln\frac{\bmu}{T}
    + \frac{24}{11} \ln A%
    - \frac{4}{11}\times\bigl[-0.299378\bigr]
  \nn &\hphantom{{}=
    \frac{g^2 \Nc T^2}{96\pi^2}\frac{11}{3}\bigg[
        \frac{1}{\epsilon}
      + 4 \ln\frac{\bmu}{T}}
      - \frac{12}{11} \gammaE
      + \frac{61}{33}
      + \frac{10}{11} \ln\pi
      - \frac{52}{11} \ln 2
    \bigg]
  \,,\\[2mm]
\label{zcexpreuclmasterfermion}
    Z^{(c)}_{\mathrm{g},q^{-2},\rmii{F}} &=
    -\frac{g^2 \Tf\Nf T^2}{72\pi^2}\bigg[
      \frac{1}{\epsilon}
      + 4 \ln\frac{\bmu A^3}{4\pi T}
      + \frac{5}{2}\times\bigl[-0.199478\bigr]
      + 3 \gammaE
      + \frac{13}{6}
    \bigg]
  \,.
\end{align}

\subsubsection{Evaluation of $J_3$ and $J_5$ and their fermionic counterparts}
\label{sub:eucl_integrals}

For the
bosonic $J_3$ and
fermionic $\hat{J}_3$, we first perform the Matsubara sums and drop
the scale-free vacuum $\times$ vacuum term, i.e.
\begin{align}
  J_3 &= \Tint{\QE\KE}\frac{1}{\QE^2\KE^2(\vec{q}-\vec{k})^2} =
  \int_{\vec{q}\vec{k}}\! \frac{[1+2 \nB(q)][1+2\nB(k)]}
    {4 qk(\vec{q}-\vec{k})^2}=
  \int_{\vec{q}\vec{k}}\! \frac{\nB(q)[1+\nB(k)]}
    { qk(\vec{q}-\vec{k})^2}
    \,, \\
    \hat J_3 &= \Tint{\{\QE\KE\}}\frac{1}{\QE^2\KE^2(\vec{q}-\vec{k})^2}
    =-\int_{\vec{q}\vec{k}}\! \frac{\nF(q)[1- \nF(k)]}
    { qk(\vec{q}-\vec{k})^2}
    \,,
\end{align}
where we have used the $\vec{q}\leftrightarrow \vec{k}$ symmetry of the integrand.
We can now split the vacuum and thermal parts in $k$. The former can be treated as follows
\begin{align}
\label{j30start}
J_3^{(0)} &=
  +\int_{\vec{q}\vec{k}}\bigg[\frac{ \nB(q)}
    { qk(\vec{q}-\vec{k})^2}-\frac{ \nB(q)\theta(k-q)}
    { qk^3}+\frac{ \nB(q)\theta(k-q)}
    { qk^3}\bigg]
    \,,\\
\label{j30startF}
  \hat J_3^{(0)}&=
  -\int_{\vec{q}\vec{k}}\bigg[\frac{ \nF(q)}
    { qk(\vec{q}-\vec{k})^2}-\frac{ \nF(q)\theta(k-q)}
    { q k^3}+\frac{ \nF(q)\theta(k-q)}
    { q k^3}\bigg]
    \,.
\end{align}
The sum of the first two terms in both cases is now finite and can be treated in $d=3$, whereas the third term is doable in DR.
We then find
\begin{align}
\label{j30}
  J_3^{(0)}&=
  \frac{T^2}{24\pi^2}+\frac{T^2}{48\pi^2}\bigg[\frac{1}{\epsilon}-4\ln\frac{4\pi T}{\bmu A^{12}}\bigg]
  \,,\\
\label{j30F}
  \hat J_3^{(0)}&=
  -\frac{T^2}{48\pi^2}-\frac{T^2}{96\pi^2}\bigg[\frac{1}{\epsilon}-4\ln\frac{8\pi T}{\bmu A^{12}}\bigg]
  \,,
\end{align}
where the first term comes from the finite part and the second from the DR part.

We now turn to the thermal $k$
contribution.
There we need to subtract its zero mode only for the bosonic integral and treat it separately in DR,
i.e.
\begin{equation}
    J_3^{(T)}=
      \int_{\vec{q}\vec{k}}\bigg[
      \frac{ \nB(q)\nB(k)}{qk(\vec{q}-\vec{k})^2}
    - \frac{ \nB(q)T}{ qk^2(\vec{q}-\vec{k})^2}
    + \frac{ \nB(q)T}{ qk^2(\vec{q}-\vec{k})^2}
    \bigg]
    \,,\quad
    \hat J_3^{(T)}=
      \int_{\vec{q}\vec{k}}\frac{\nF(q)\nF(k)}{qk(\vec{q}-\vec{k})^2}
    \,.
\end{equation}
Once again, the sum of the first two terms is finite. It can be treated numerically in three dimensions. The third 
term is dealt with in DR, using standard vacuum techniques for the $\vec{k}$ integration followed by the $\vec{q}$ one.
We find
\begin{equation}
    J_3^{(T)}=\frac{T^2}{64\pi^2}\bigg[-3.25532 -\frac{1}{\epsilon}-4\ln\frac{\bmu}{T}-2\bigg].
\end{equation}
Using the position-space method of~\cite{Arnold:1994ps,Arnold:1994eb},
the numerical part can be integrated analytically with full
agreement, finding
\begin{align}
  J_3^{(T)} &=
      \frac{T^2}{48\pi^2}\bigg[1+3 \ln\frac{2 \pi }{A^{12}}\bigg]
    + \frac{T^2}{64\pi^2}\bigg[ -\frac{1}{\epsilon}-4\ln\frac{\bmu}{T}-2\bigg]
    \,,\\
    \hat J_3^{(T)} &= \frac{T^2}{48\pi^2}\Bigl[1-\ln 2\Bigr]
    \,.
\end{align}
Hence
\begin{align}
\label{j3final3}
      J_3=J_3^{(0)}+J_3^{(T)} &=
      \frac{T^2}{192\pi^2}\bigg[\frac{1}{\epsilon}-4\ln\frac{32\pi T}{\bmu A^{12}}+6\bigg]
      \,,\\
\label{j3ffinal}
    \hat J_3 =\hat J_3^{(0)}+\hat J_3^{(T)} &=
   -\frac{T^2}{96\pi^2}\bigg[\frac{1}{\epsilon}-4\ln\frac{8\pi T}{\bmu A^{12}}+\ln 4
    \bigg]
    \,.
\end{align}

To evaluate $J_5$,
we employ again the positon-space methods of~\cite{Arnold:1994ps} and
write $J_5$ as
\begin{equation}
    J_5=\Tint{\QE\KE}\frac{k_0^2}{\QE^2q^2\KE^2(\QE+\KE)^2}=\Tint{\QE}\frac{1}{\QE^2q^2}\Pi(\QE)
    \,,\quad
    \label{defj5AZ}
\end{equation}
where
we can separate the vacuum and thermal contribution to $\Pi(Q)$
\begin{align}
\label{pisplit}
    \Pi(\QE) &\equiv\Tint{\KE}\frac{k_0^2}{\KE^2(\QE+\KE)^2}
    =
      \Pi^{(0)}(\QE)
    + \Pi^{(T)}(\QE)
    \,.
\end{align}
The vacuum contribution is obtainable by replacing the Matsubara sum over $k_0$ with a Euclidean
$\int {\rm d}k_0/(2\pi)$.
Employing standard DR methods and 
inserting the result into Eq.~\eqref{defj5AZ}, yields
\begin{equation}
    J_5^{(0)}=\mu^{6-2d} \frac{ ((d-2) d-1)  T^{2 d-4}}{(4\pi)^{d+1}\pi^{\frac{7}{2}-2d}}
   \frac{
     \Gamma \bigl(\frac{5}{2}-d\bigr)
     \Gamma \bigl(d-3\bigr)}{
     \Gamma \bigl(\frac{d}{2}\bigr)
     \Gamma \bigl(\frac{d+2}{2}\bigr)}\zeta_{5-2 d}=
   -\frac{T^2}{576\pi^2}\bigg[
   \frac{1}{\epsilon}+4 \ln\frac{\bmu A^{12}}{4\pi T}+\frac{2}{3}\bigg]
   \,.
   \label{j5vac}
\end{equation}
For $\Pi^{(T)}$, we instead proceed in position space, i.e.\
\begin{align}
    \Pi(\QE)&=\Tint{\KE}\frac{k_0^2}{\KE^2(\QE+\KE)^2}=
    \int\!{\rm d}^3\vec{r}\,\Tint{\KE\PE}\!\frac{k_0^2\beta\delta_{p_0-k_0-q_0}}{\KE^2\PE^2}
    e^{i\vec{r}\cdot(\vec{p}+\vec{q}+\vec{k})}
    \nn[1mm] &=
    T\sum_{k_0}\int\!{\rm d}^3\vec{r}
    \frac{k_0^2e^{-\vert k_0\vert r}e^{-\vert q_0+k_0\vert r}}{(4\pi r)^2}e^{i\vec{q}\cdot\vec{r}}
    \,,
\end{align}
where $\delta_{p_0} = \delta_{p_0,0}$ and
in the last step we took
the three-dimensional Fourier transforms.
One can also work out the Matsubara sum, finding
 \begin{equation}
     \Pi(\QE)=T^3\int\!{\rm d}^3\vec{r}\frac{e^{-\vert\bar{q}_0\vert \bar{r}}e^{i\vec{q}\cdot\vec{r}}}{12 r^2}\bigg[ \vert\bar{q}_0\vert^3
     + \frac{3}{2}\Bigl(\bar{q}_0^2 \coth (\bar{r})+\mbox{csch}^2(\bar{r})
        \bigl(\vert\bar{q}_0\vert+ \coth (\bar{r})\bigr)\Bigr)
    + \frac{\vert\bar{q}_0\vert}{2}
  \bigg]
   \,,
\end{equation}
where $\bar{q}_0\equiv q_0/(2\pi T)$ and $\bar{r}\equiv 2\pi Tr$.
To get $\Pi^{(T)}$, we must subtract the vacuum contribution, again
obtained by replacing the sum with the integral.
We then find
\begin{align}
  \Pi^{(T)}(\QE)=T^3\int\!{\rm d}^3\vec{r}\frac{e^{-\vert\bar{q}_0\vert \bar{r}}e^{i\vec{q}\cdot\vec{r}}}{24 r^2}
    \bigg[
      3\biggl(&
        \bar{q}_0^2\Bigl( \coth (\bar{r})-\frac{1}{\bar r}\Bigr)
      + \vert\bar{q}_0\vert\Bigl(\text{csch}^2(\bar{r})-\frac{1}{\bar r^2}\Bigr)
   \nn &
      + \text{csch}^2(\bar{r})\coth (\bar{r})-\frac{1}{\bar r^3}
    \biggr)
    + \vert\bar{q}_0\vert 
  \bigg].
  \label{piTrspace}
\end{align}
In principle,
we should naively insert this into Eq.~\eqref{defj5AZ}.
We would find that
\begin{itemize}
  \item[1)]
    Since the Fourier transform 
    \begin{equation}
        \int_\vec{q}\frac{e^{i\vec{q}\cdot\vec{r}}}{q^2 \QE^2}=\frac{1-e^{-\vert q_0\vert r}}{4\pi q_0^2 r}
        (1-\delta_{q_0})
        \,,
        \label{qtransform}
    \end{equation}
    is only defined in 3D for the non-zero modes, we need to treat the $\QE$ zero mode separately.
  \item[2)] If we use Eqs.~\eqref{piTrspace} and \eqref{qtransform} together, perform the nonzero 
    $q_0$ sum, we are then faced with a ${\rm d}r/r$ UV logarithmic divergence. 
    As per \cite{Arnold:1994ps}, we must then subtract off the $\QE\gg T$ limit of Eq.~\eqref{piTrspace}
    in position space and add it back in momentum space in DR.
\end{itemize}
We then find
 \begin{align}
     \Pi^{(T)}(\QE\gg T)=&T^3\int\!{\rm d}^3\vec{r}\frac{e^{-\vert\bar{q}_0\vert \bar{r}}e^{i\vec{q}\cdot\vec{r}}}{24 r^2 }\bar{q}_0^2\bar{r}+\ldots,
  \label{piTrspaceUV}
\end{align}
where the dots stand for higher-order terms in the expansion for $\bar r\ll 1$ with $\bar r \bar{q}_0\sim 1$.
In momentum space the analogue expansion can be made by having either of the $\KE$ and $\QE+\KE$ denominators 
in Eq.~\eqref{defj5AZ} be much larger than $T$. The leading term comes from the former, yielding
 \begin{align}
     \Pi^{(T)}(\QE\gg T)=\frac{q_0^2}{\QE^2}\Tint{\KE}\frac{1}{\KE^2}+\mathcal{O}\left(\frac{T^4}{\QE^2}\right).
  \label{piTQspaceUV}
\end{align}

By splitting $J_5^{(T)}$ into three terms and
noting that primed sum-integrals exclude the zero mode,
we obtain
\begin{align}
  J_5^{(T)}&=
        \Tint{\QE}'\bigg[\frac{\Pi^{(T)}(\QE)}{\QE^2q^2}-\frac{\Pi^{(T)}(\QE\gg T)}{\QE^2q^2}\bigg]
      + \Tint{\QE}'\frac{q_0^2}{\QE^4q^2}\Tint{\KE}\frac{1}{\KE^2}
      + \Tint{\QE\KE}\!\frac{\delta_{q_0}k_0^2}{q^4 \KE^2(\QE+\KE)^2}
    \nn[2mm] &=
        J_5^{(T)\,\text{finite}}
      + J_5^{(T)\,\text{div.}}
      + J_5^{(T),q_0=0}
    \label{defj5split}
    \,.
\end{align}

For the first term we proceed in position space.
After using Eq.~\eqref{qtransform}
and performing the $q_0$ sum, we have
\begin{align}
J_5^{(T)\,\text{finite}}&=
    T^4\sum_{q_0\ne0}
    \int{\rm d}^3\vec{r}
    \frac{(1-e^{-\vert \bar q_0\vert \bar{r}})e^{-\vert\bar{q}_0\vert \bar{r}}}{4\pi q_0^2 r}
    \frac{1}{8 r^2 }\biggl[
      \Bigl(\mbox{csch}^2(\bar{r})\coth (\bar{r})-\frac{1}{\bar r^3}\Bigr)
   \nn &\hphantom{{}=T^4\sum_{q_0\ne0} \int{\rm d}^3\vec{r}}
      + \vert\bar{q}_0\vert^{ }\Bigl(\mbox{csch}^2(\bar{r})-\frac{1}{\bar r^2} + \frac{1}{3} \Bigr)
      + \vert\bar{q}_0\vert^{2}\Bigl( \coth (\bar{r})-\frac{1}{\bar r}-\frac{\bar{r}}{3}\Bigr)
  \biggr]
  \nn[2mm] &=\frac{T^5}{8}
  \int{\rm d}^3\vec{r}\biggl\{%
   \Bigl[
        \mbox{Li}_2\left(e^{-\bar{r}}\right)
      - \mbox{Li}_2\left(e^{-2 \bar{r}}\right)\Bigr]
   \Bigl(
      \frac{\coth(\bar{r}) \mbox{csch}^2(\bar{r})}{ \bar{r}^3}
    - \frac{1}{\bar{r}^6}\Bigr)
   \nn &\hphantom{{}=\frac{T^5}{8} \int{\rm d}^3\vec{r}}
   + \ln\left(e^{-\bar{r}}+1\right) 
   \Bigl(
        \frac{\mbox{csch}^2(\bar{r})}{\bar{r}^3}
      + \frac{1}{3\bar{r}^3}
      -\frac{1}{ \bar{r}^5}\Bigr)
    + \frac{\mbox{csch}(\bar r)}{6 \bar{r}^2}\Bigl(\frac{3\coth(\bar r)}{\bar r}-1-\frac{3}{\bar r^2}\Bigr)
   \biggr\}
   \nn[2mm]
   &= \bigl[-0.0997928 \bigr] \times \frac{T^2}{96\pi^2}
   \,,
   \label{j5num}
\end{align}
where the integration is finite and has been carried out numerically.%
\footnote{An analytical evaluation could be 
possible by using IBP:
$\ln(1+e^{-\bar r})=\ln (1-e^{-2\bar r})-\ln(1-e^{-\bar{r}})$, and
these two terms are related to the derivatives of the polylogarithms.
The non-logarithmic and non-polylogarithmic 
term can be done using the methods of~\cite{Arnold:1994ps,Arnold:1994eb}, giving
\begin{equation}
  \frac{T^5}{8} \int{\rm d}^3\vec{r}\frac{\text{csch}(\bar r)}{6 \bar{r}^2}\bigg(\frac{3\coth(\bar r)}{\bar r}-1-\frac{3}{\bar r^2}\bigg)=\frac{T^2}{96 \pi ^2}\bigg[ \ln\frac{A^{12}}{8}-\frac12- \gammaE \bigg]
  \,.
  \nonumber
\end{equation}
}

For the divergent piece we find, using
IBP 
\begin{align}
\label{j5div}
  J_5^{(T)\,\rmi{div.}} &= \Tint{\QE}' \frac{q_0^2}{\QE^4 q^2}
    \Tint{\KE} \frac{1}{\KE^2} =
    \Bigl[
    \frac{4-d}{2}\Tint{\QE} \frac{1}{\QE^2 q^2}
    \Bigr]
    \mathcal{I}_{1}
    =
    -\frac{d-4}{d-2}\mathcal{I}_{1}\mathcal{I}_{2}
    \nn &=
   \frac{T^2}{192\pi^2}\bigg[\frac{1}{\epsilon}
      + 4 \ln\frac{\bmu A^{6}}{4\pi T}
      + 2\gammaE + 4
   \bigg]\,.
\end{align}
Finally, the zero mode can be computed
using Feynman parametrisation and
the series expansion for the Riemann zeta function.
Below,
we list the bosonic~\cite{Moller:2010xw} and fermionic results
\begin{align}
\label{zeromodemasterboson}
  \mathcal{Z}_{s_1 s_2 s_3}^{\alpha} &=
  \Tint{\QE\KE}
  \!\delta_{q_0}
  \frac{(k_0^{2})^{\alpha}}{
    [\QE^2]^{s_1}
    [\KE^2]^{s_2}
    [(\QE-\KE)^2]^{s_3}}
  \nn
  &=
  \mu^{6-2d}
  \frac{
    \Gamma\bigl( \frac{d}{2} - s_1\bigr)
    \Gamma\bigl( s_{12} - \frac{d}{2}\bigr)
    \Gamma\bigl( s_{13} - \frac{d}{2}\bigr)
    \Gamma\bigl( s_{123} - d \bigr)}{
    (4\pi)^d
    \Gamma\bigl( \frac{d}{2} \bigr)
    \Gamma\bigl( s_2 \bigr)
    \Gamma\bigl( s_3 \bigr)
    \Gamma\bigl( s_{1123} - d \bigr)
  }
  \frac{
    2T^{2} \zeta_{2s_{123} - 2\alpha - 2d}}{
    (2\pi T)^{2s_{123} - 2\alpha - 2d}}
  \,,\\
  \widehat{\mathcal{Z}}_{s_1 s_2 s_3}^{\alpha} &=
  \Tint{\QE\{\KE\}}
  \!\delta_{q_0}
  \frac{(k_0^{2})^{\alpha}}{
    [\QE^2]^{s_1}
    [\KE^2]^{s_2}
    [(\QE-\KE)^2]^{s_3}}
  =
  (2^{2s_{123} - 2\alpha - 2d} -1) \mathcal{Z}_{s_1 s_2 s_3}^{\alpha}
    \,,
\end{align}
where
$s_{\{i\}} = \sum_{j\in \{i\}} s_j$.
The two special cases relevant for our computation
\begin{align}
\label{j5zeromode}
  \mathcal{Z}_{211}^{1} &\stackrel{d=3-2\epsilon}{=}
    J_5^{(T)\;q_0=0}
    =
    \frac{T^2}{128\pi^2}
  \,,\\
\label{j5zeromodeF}
  \widehat{\mathcal{Z}}_{211}^{1} &\stackrel{d=3-2\epsilon}{=}
    \hat J_5^{(T)\;q_0=0}
    =0
    \,,
\end{align}
can also be
read from Eqs.~(D2) and (D5) of \cite{Brambilla:2010xn}.
Summing up Eqs.~\eqref{j5vac}, \eqref{j5num}, \eqref{j5div} and \eqref{j5zeromode} we find
Eq.~\eqref{eq:J5}.

For $\hat{J}_5$ we adopt the same strategy used for its bosonic counterpart.
The $\KE$ vacuum part is identical and given by Eq.~\eqref{j5vac}. 
The $\KE$ thermal part is split in three terms (cf.\ Eq.~\eqref{defj5split})
\begin{align}
  \hat J_5^{(T)} &=
      \hat J_5^{(T)\,\rmi{finite}}
    + \hat J_5^{(T)\,\rmi{div.}}
    + \hat J_5^{(T),q_0=0}
    \,.
\end{align}
The finite and divergent terms are
\begin{align}
  \hat J_5^{(T)\,\rmi{finite}}&=
  \frac{T^5}{8} \int{\rm d}^3\vec{r}\biggl\{
   \Bigl[
        \mbox{Li}_2\left(e^{-\bar{r}}\right)
      - \mbox{Li}_2\left(e^{-2 \bar{r}}\right)\Bigr]
   \Bigl(\frac{\mbox{csch}^3(\bar r)(\cosh (2 \bar r)+3)}{4\bar r^3}-\frac{1}{\bar r^6}\Bigr)
   \nn &\hphantom{{}=\frac{T^5}{8} \int{\rm d}^3\vec{r}}
    + \ln\left(e^{-\bar{r}}+1\right)
      \Bigl(\frac{6 \coth(\bar r)\,\text{csch}(\bar{r})-1}{6\bar{r}^3}-\frac{1}{ \bar{r}^5}\Bigr)
   \nn &\hphantom{{}=\frac{T^5}{8} \int{\rm d}^3\vec{r}}
    + \frac{\text{csch}(\bar r)}{12\bar{r}^2}\Bigr(\frac{6\,\text{csch}(\bar r)}{\bar r}+1-\frac{6}{\bar r^2}\Bigr)
    \bigg\}
  \nn[2mm] &=
  \bigl [ 0.0831159 \bigr] \times \frac{T^2}{96\pi^2}
  \label{j5numF}
  \,,\\[2mm]
\label{j5divF}
  \hat J_5^{(T)\,\rmi{div.}} &=
  (2^{2-d}-1) J_5^{(T)\,\rmi{div.}} \stackrel{\text{\nr{j5div}}}{=}
  -\frac{T^2}{384\pi^2}\bigg[
        \frac{1}{\epsilon}
      + 4\ln\frac{\bmu A^{6}}{4\pi T}
      + 2\gammaE +4-\ln4
   \bigg]
   \,.
\end{align}
Finally, the zero mode can be read off from Eq.~\eqref{j5zeromodeF}.
Summing up Eqs.~\eqref{j5vac}, \eqref{j5numF}, \eqref{j5divF} and \eqref{j5zeromodeF} we find
Eq.~\eqref{eq:J5F}.

\subsection{Real-time evaluation of $Z_{\mathrm{g},\rmii{$L$-$T$}}$}
\label{app:realtime}
For the sake of readability,
we split our evaluation into 
the fermionic quark- and
the bosonic gauge-contribution.

\subsubsection{Quark contribution}\label{app:fermion_lmt}

The unresummed Hard Thermal Loop (HTL) subset of
the full contribution, using Eq.~\eqref{retdiffquarksimple},
is
\begin{align}
  Z_{\mathrm{g},\rmii{$L$-$T$},\rmii{HTL},\rmii{F}}^{(c)} &=
  i\int_{Q}\frac{1+\nB(q^0)}{(q^--i\varepsilon)^2}\bigg[
 \frac{q_\perp^2(\Pi_\rmii{$L$}^{\rmii{$R$}}(Q)-\Pi_\rmii{$T$}^{\rmii{$R$}}(Q))_\rmii{HTL}}{q^2(Q^2+i\varepsilon q^0)}-\text{adv.}
    \bigg]\,,
    \label{zcexprLTHTL}
    \\[2mm]
   \Pi_\rmii{$L$}^{\rmii{$R$}}(Q)_{\rmii{F},\rmii{HTL}}-\Pi_\rmii{$T$}^{\rmii{$R$}}(Q)_{\rmii{F},\rmii{HTL}} &= 
   \frac{g^2\Tf\Nf T^2}{12}
   \bigg[-  \frac{(6 q_0^2-4q^2)}{q^2}
  +3 \frac{ q^0Q^2}{q^3} \ln \frac{q^0+q+i\varepsilon }{q^0-q+i\varepsilon} 
   \bigg]\,,
   \label{retdiffquarksimpleHTL}
\end{align}
which, after inserting
Eq.~\eqref{retdiffquarksimpleHTL}
into~\eqref{zcexprLTHTL},
gives rise to a pole and a cut contribution.
The pole contribution is
\begin{align}
 \label{zcexprLTHTLpole}
  Z_{\mathrm{g},\rmii{$L$-$T$},\rmii{F},\rmii{HTL}\,\rmi{pole}}^{(c)} &=
  -\frac{g^2 \Tf \Nf T^2}{12}
    \int_{Q}\frac{\frac{1}{2}+\nB(\vert q^0\vert)}{(q^- -i\varepsilon)^2}\frac{(6 q_0^2-4q^2)}{q^2}
 \frac{q_\perp^2}{q^2}2\pi\delta(Q^2)
    \\[1mm]
\label{zcexprLTHTLpoleDRfinal}
    &=
  -\frac{g^2 \Tf \Nf T^2}{6}
    \int_{\vec{q}}\frac{\nB(q)}{2q}\bigg[\frac{1}{(q-q^z-i\varepsilon)^2}
    +\frac{1}{(-q-q^z-i\varepsilon)^2}\bigg] \frac{q_\perp^2}{q^2}
    \\ &=
    \frac{g^2 \Tf\Nf T^2}{48\pi^2}\bigg[\frac{1}{\epsilon^2}
    + \frac{2}{\epsilon} \ln\frac{\bmu e^{\gammaE+\frac12}}{4\pi T}
    + 2 \ln^2\frac{\bmu e^{\gammaE+\frac12}}{4\pi T}
    -4 \gamma _1+\frac{\pi^2}{4}
    -2 \gammaE^2+\frac{3}{2}
    \bigg]
    \,,
    \nonumber
\end{align}
where in DR, the second step used that
the vacuum part is scale-free.
The cut contribution of the HTL piece is instead
\begin{align}
  Z_{\mathrm{g},\rmii{$L$-$T$},\rmii{F},\rmii{HTL}\,\rmi{cut}}^{(c)}&=
  \frac{g^2 \Tf\Nf T^2}{4}
    \int_{\vec{q}}  \int_0^q \frac{{\rm d}q^0}{2\pi}\frac{q_\perp^2 q^0}{q^5}2\pi %
     \bigg[
        \frac{1+\nB(q^0)}{(q^0-q^z-i\varepsilon)^2}
      + \frac{\nB(q^0)}{(q^0+q^z+i\varepsilon)^2}
     \bigg]
   \,,
     \label{zcexprLTHTLcutsympos}
\end{align}
where we have symmetrised the frequency integral before restricting the integration to be over twice the positive range. The $\nB$-independent part is now scale-free, while the rest can be done
in a somewhat similar way to the pole piece
\begin{align}
  Z_{\mathrm{g}\,\rmii{$L$-$T$}\,\rmii{F}\,\rmii{HTL}\,\rmi{cut}}^{(c)}&=
  \frac{g^2 \Tf \Nf T^2}{4}\Big(\frac{\bmu^{2}e^{\gammaE}}{4\pi} \Big)^\epsilon
  \frac{2\pi^{\frac{d-1}{2}}}{\Gamma(\frac{d-1}{2})(2\pi)^{d-1}}
    \int_0^\infty \frac{{\rm d}q^0}{2\pi} q^0 \nB(q^0)
    \nn &\times
    \int_{q_0}^\infty\!{\rm d}q q^{d-4} \int_{-1}^{1} {\rm d}x (1-x^2)^{\frac{d-1}{2}}
   \bigg[
    \frac{1}{(q^0-q x-i\varepsilon)^2}+
    \frac{1}{(q^0+q x+i\varepsilon)^2}\bigg]
  \,.
    \label{zcexprLTHTLcutsymposDR}
\end{align}
Our strategy here is to do the following subtraction:
$
\nB(q^0)\to
\nB(q^0)\mp(T/q^0-\theta(T-q^0)/2)
$.
In this way, the subtraction
is finite, and then we re-evaluate the addition in DR. 
For the finite part, we find, performing the angular and momentum integrations first
\begin{align}
  Z_{\mathrm{g},\rmii{$L$-$T$},\rmii{F},\rmii{HTL}\,\rmi{cut finite}}^{(c)}&=
  \frac{g^2 \Tf \Nf T^2}{4} \frac{1}{2\pi}
    \int_0^{\infty} \frac{{\rm d}q^0}{2\pi} q^0 \bigg[\nB(q^0)-\frac{T}{q^0}
    +\frac{\theta(T-q^0)}{2}\bigg] 
    \nn &\times
    \int_{q^0}^\infty \frac{{\rm d}q}{q}\int_{-1}^{1}\!{\rm d}x (1-x^2)
     \bigg[
    \frac{1}{(q^0-q x-i\varepsilon)^2}+
    \frac{1}{(q^0+q x+i\varepsilon)^2}\bigg]
    \nn[2mm] &=
    \frac{g^2 \Tf \Nf T^2}{12 \pi^2}\ln\frac{e^{\gammaE}}{2\pi}(\ln2-1)
    \,.
        \label{zcexprLTHTLcutsymposDRfinite}
\end{align}
For the DR subtracted part,
we instead do the frequency integral first, splitting it into the
$q>T$ and
$q<T$ ranges,
\begin{align}
  Z_{\mathrm{g},\rmii{$L$-$T$},\rmii{F},\rmii{HTL}\,\rmi{cut div}}^{(c)}&=
  -\frac{g^2 \Tf \Nf T^2}{8}\Big(\frac{\bmu^{2}e^{\gammaE}}{4\pi} \Big)^\epsilon
  \frac{2\pi^{\frac{d-1}{2}}}{\Gamma \left( \frac{d-1}{2} \right)(2\pi)^{d}}
  \int_{0}^\infty\! {\rm d}q \, q^{d-4} \int_{-1}^{1}\! {\rm d}x \, (1-x^2)^{\frac{d-1}{2}}
  \nn & \times
    \bigg\{\theta(T-q)\bigg[-\mathbb{P}\frac{2}{1-x^2}+\ln\frac{1-x^2}{x^2}\bigg]
  \nn &\hphantom{{}\biggl\{}
  +\theta(q-T)\bigg[\ln\frac{\left\vert T^2-q^2 x^2\right\vert }{q^2 x^2}-\mathbb{P}\frac{2 T^2}{T^2-q^2 x^2}\bigg]
   \bigg\}
   \,,
     \label{zcexprLTHTLcutsymposDRdivqzero}
\end{align}
where the principal value (PV) $\mathbb{P}$ arises 
from the $\varepsilon\to 0$ limit.
The $q>T$ slice is finite whereas the
$T>q$ part is IR-divergent and needs regularisation
\begin{align}
  Z_{\mathrm{g},\rmii{$L$-$T$},\rmii{F},\rmii{HTL}\,\rmi{cut div}}^{(c)}&=
  -\frac{g^2 \Tf \Nf T^2}{8}\Big(\frac{\bmu^{2}e^{\gammaE}}{4\pi} \Big)^\epsilon
  \frac{2\pi^{\frac{d-1}{2}}}{\Gamma \left( \frac{d-1}{2} \right) (2\pi)^{d}}
  \nn &
    \times\bigg\{
      \int_{-1}^{1} {\rm d}x \, (1-x^2)^{\frac{d-1}{2}}
      \frac{T^{d-3}}{d-3}\bigg[-\mathbb{P}\frac{2}{1-x^2}+\ln\frac{1-x^2}{x^2}\bigg]
    \nn &\hphantom{{}\biggl\{}
    + \int_{-1}^{1} {\rm d}x \, (1-x^2)\bigg[\frac{\mbox{Li}_2\left(x^2\right)}{2}+\ln \left(\frac{x^2}{1-x^2}\right)+\ln ^2\vert x\vert-\frac{\pi ^2}{6}\bigg]
   \bigg\}
   \,,
     \label{zcexprLTHTLcutsymposDRdivqzerosplit}
  \\[2mm] &=
  -\frac{g^2 \Tf \Nf T^2}{8}\Big(\frac{\bmu^{2}e^{\gammaE}}{4\pi} \Big)^\epsilon
  \frac{2\pi^{\frac{d-1}{2}}}{\Gamma \left( \frac{d-1}{2} \right) (2\pi)^{d}}
  \nn &\times \bigg\{%
    \frac{T^{d-3}}{d-3}
    \frac{
        \Gamma \bigl(\frac12\bigr)
        \Gamma \bigl(\frac{d+1}{2}\bigr)}{\Gamma \bigl(\frac{d}{2}+1\bigr)}
        \Bigl(
            \psi\Bigl(\frac{d-1}{2}\Bigr)
          + \gammaE 
          - 2
          + 2\ln 2
          \Bigr)
    -\frac{1}{9} \left(8+\pi^2-8\ln 2\right)
   \bigg\}
  \nn[2mm] &=
  \frac{g^2 \Tf \Nf T^2}{24\pi^2}(\ln 2-1)\bigg[
      \frac{1}{\epsilon}
    + 2\ln\frac{\bmu}{2T}+1 %
   \bigg]
  \,,
  \label{zcexprLTHTLcutsymposDRdivqzerosplitfinal}
\end{align}
where $\psi(x) = (\ln\Gamma(x))'$ is the digamma function.
Finally
\begin{equation}
  Z_{\mathrm{g},\rmii{$L$-$T$},\rmii{F},\rmii{HTL}\,\rmi{cut}}^{(c)} = 
    \frac{g^2 \Tf \Nf T^2}{24\pi^2}(\ln2-1)\bigg[
      \frac{1}{\epsilon}+2\ln\frac{\bmu e^{\gammaE+\frac12}}{4\pi T}\bigg]
    \,.
   \label{zcexprLTHTLcutfinal}
\end{equation}
Let us now consider the non-HTL piece
\begin{align}
     \delta\Pi_{\rmii{$L$}}^{\rmii{$R$}}(Q)_\rmii{F}
   - \delta\Pi_{\rmii{$T$}}^{\rmii{$R$}}(Q)_\rmii{F} =
   \frac{2g^2 \Tf \Nf Q^2}{(4\pi)^2 q^3}
   &
   \int_0^\infty\!{\rm d}k \, \nF(k)  \bigg[24  k q
   \nn &
   + \left(3 (2k-q^0)^2-q^2\right) \ln\frac{q^0+q-2k+i\varepsilon}{ q^0-q-2k+i\varepsilon}
   \nn &
   + \left(3 (2k+q^0)^2-q^2\right)\ln\frac{q^0-q+2k+i\varepsilon}{ q^0+q+2k+i\varepsilon}
   \bigg]
   \,,
   \label{retdiffquarksimplenonHTL}
\end{align}
where $\delta\Pi=\Pi-\Pi_\rmii{HTL}$. 
This then gives
\begin{align}
  Z_{\mathrm{g},\rmii{$L$-$T$},\rmii{F},\rmii{non-HTL}}^{(c)}&=
    \frac{g^2 \Tf \Nf }{4\pi}\int_0^\infty\!{\rm d}k \, \nF(k) \int_{Q}\frac{\frac{1}{2}+\nB(q^0)}{(q^--i\varepsilon)^2}
  \frac{q_\perp^2}{q^5}
  \nn &\times
  \bigg[
      \theta(q^2-(q^0 - 2 k)^2)\left(3(2k-q^0)^2-q^2\right)
  \nn &\hphantom{\Bigl[}
    - \theta(q^2-(q^0 + 2 k)^2) \left(3(2k+q^0)^2-q^2\right)
  \bigg]
  \,.
    \label{zcexprlogs}
\end{align}
Here, we used the fact that the $\frac12$ piece vanishes%
\footnote{%
  It gives rise to fully retarded 
  and fully advanced contributions in $q^+$.
  The contours can be safely closed in these half-planes 
  without encountering singularities or contributions from the arcs at large $\vert q^+\vert$,
  thus yielding a vanishing contribution.
} to restrict to the purely odd 
$\frac12+\nB(q^0)$ component.
Let us again symmetrise and restrict to twice the positive frequency range
\begin{align}
  Z_{\mathrm{g},\rmii{$L$-$T$},\rmii{F},\rmii{non-HTL}}^{(c)}&=\frac{g^2 \Tf \Nf }{4\pi}\int_0^\infty\!{\rm d}k \, \nF(k) \int_0^\infty\frac{{\rm d}q^0}{2\pi}
    \int_{\vec{q}}\frac{q_\perp^2}{q^5}\bigg[\frac{\frac12+\nB(q^0)}{(q^0-q^z-i\varepsilon)^2}
    +\frac{\frac12+\nB(q^0)}{(q^0+q^z+i\varepsilon)^2}\bigg]\nn
    &\times
  \bigg\{
    \theta(q-2k)\bigg[
      \left (3(2k-q^0)^2-q^2\right)\theta(2k+q-q^0)
  \nn &\hphantom{\times\bigg\{\theta(q-2k)}
    - \left (3(2k+q^0)^2-q^2\right)\theta(q-2k-q^0)\bigg]
  \nn &
  +\theta(2k-q) \left (3(2k-q^0)^2-q^2\right)\theta(2k+q-q^0)
  \theta(q^0+q-2k)\bigg\}
  \,.
    \label{zcexprlogssymthetas}
\end{align}
We start by dealing with the vacuum part
\begin{align}
  Z_{\mathrm{g},\rmii{$L$-$T$},\rmii{F},\rmii{non-HTL}\,\rmi{vac}}^{(c)}&=
    \frac{g^2 \Tf\Nf}{8 \pi}\int_0^\infty\!{\rm d}k \, \nF(k) \int_0^\infty\frac{{\rm d}q^0}{2\pi}
    \int_{\vec{q}}\frac{q_\perp^2}{q^5}\bigg[\frac{1}{(q^0-q^z-i\varepsilon)^2}
    +\frac{1}{(q^0+q^z+i\varepsilon)^2}\bigg]
  \nn& \times
  \bigg\{\theta(q-2k)\bigg[\left (3(2k-q^0)^2-q^2\right)\theta(q+2k-q^0)
  \nn &\hphantom{\times\bigg\{\theta(q-2k)}
  -\left (3(2k+q^0)^2-q^2\right)\theta(q-2k-q^0)\bigg]
  \nn &
  + \theta(2k-q)\left (3(2k-q^0)^2-q^2\right)\theta(2k+q-q^0)\theta(q^0-2k+q)
  \bigg\}
  \,.
    \label{zcexprlogssymvac}
\end{align}
The strategy here is the following: we further symmetrise the expression in $q^z$,
to perform the $q^0$ integrations. After that one can take $\varepsilon\to 0$.
Logarithmic pieces will see their argument change to its absolute value, while
possible imaginary parts will cancel out due to $q^z$ symmetrisation. 
Non-logarithmic pieces
present poles on the $q^z$ integration range. If we restrict to twice
the positive $q^z$ range, these occur at $-\vert 2k-q\vert+q^z=0$ and 
are turned into a PV prescription by the $\varepsilon\to 0$ limit. 
We can thus proceed to the $q^z$ integral in the two ranges $2k >q$ and $q>2k$. 
In the latter case, we find two subranges, one for $q_\perp<2k$ and one for 
$q_\perp>2k$. This last one has to be carried out in DR, since it contains a UV divergence.
We then find
\begin{align}
  Z_{\mathrm{g},\rmii{$L$-$T$},\rmii{F},\rmii{non-HTL}\,\rmi{vac}}^{(c)}=
  \frac{g^2 \Tf \Nf T^2}{48\pi^2}\bigg\{&
      \frac{1}{\epsilon^2}
    + \frac{2}{\epsilon}\ln\frac{\bmu A^{12}e^{-\frac12}}{16\pi T}
    + 2 \ln^2\frac{\bmu A^{12}e^{-\frac12}}{16\pi T}
    - \frac{7}{2}\ln^2(2)
    - 2(\ln\zeta_2)'^{2}
    \nn &
    +\frac{2\zeta_2''}{\zeta_2}
    +\frac{1}{2}\mbox{Li}_2\Bigl(\frac{2}{3}\Bigr)
    +\frac{1}{4}\mbox{Li}_2\Bigl(\frac{3}{4}\Bigr)
    -\frac12
    -\frac14 \ln(3) \ln\frac{4}{3}
    \bigg\}
    \,.
      \label{zcexprlogssymvacsympfinal}
\end{align}
For the thermal part,
we have instead
\begin{align}
  Z_{\mathrm{g},\rmii{$L$-$T$},\rmii{F},\rmii{non-HTL}\,\rmi{th}}^{(c)}&=
    \frac{g^2 \Tf \Nf }{2 \pi}\int_0^\infty\!{\rm d}k \, \nF(k) \int_0^\infty\frac{{\rm d}q^0}{2\pi}
    \int_{0}^\infty \frac{{\rm d}q}{4\pi^2}\frac{\nB(q^0)}{q}
    \bigg[
      \frac{2 q^0 \tanh ^{-1}\left(\frac{2 q q^0}{q^2+q_0^2}\right)}{q^3}
      -\frac{4}{q^2}\bigg]
  \nn & \times
  \bigg\{\theta(q-2k)\bigg[\left (3(2k-q^0)^2-q^2\right)\theta(2k+q-q^0)
  \nn &\hphantom{\times\bigg\{\theta(q-2k)}
  -\left (3(2k+q^0)^2-q^2\right)\theta(q-2k-q^0)\bigg]
  \nn &
  +\theta(2k-q) \left (3(2k-q^0)^2-q^2\right)\theta(2k+q-q^0)\theta(q^0+q-2k)\bigg\}
  \,.
    \label{zcexprlogssymthetas3Dthermal}
\end{align}
Here,
we first perform the $q$ integration analytically,
followed by $q^0$ and $k$
numerically.
We find
\begin{equation}
  Z_{\mathrm{g},\rmii{$L$-$T$},\rmii{F},\rmii{non-HTL}\,\rmi{th}}^{(c)}=
  \frac{g^2 \Tf \Nf T^2}{48\pi^2}\bigg[
      \underbrace{-0.438129\vphantom{X_{X_X}}}_{q<2k}
    + \underbrace{1.97382\vphantom{X_{X_X}}}_{q>2k}\bigg]
  \,.
\end{equation}

Putting everything together,
we find
\begin{align}
  Z_{\mathrm{g},\rmii{$L$-$T$},\rmii{F}}^{(c)}&=\frac{g^2 \Tf\Nf T^2}{48\pi^2}
      \bigg\{\frac{2}{\epsilon^2}
    + \frac{2}{\epsilon}\Bigl(
        \ln\frac{\bmu e^{\gammaE+\frac12}}{4\pi T}
      + \ln\frac{\bmu A^{12}e^{-\frac12}}{16 \pi T}
      + \ln2-1
      \Bigr)
  \nn &
    + 2 \ln^2\frac{\bmu e^{\gammaE+\frac12}}{4\pi T}
    - 4 \gamma _1+\frac{\pi ^2}{4}
    - 2 \gammaE^2+\frac32+2(\ln2-1)\bigg[
      2\ln\frac{\bmu e^{\gammaE+\frac12}}{4\pi T}\bigg]
  \nn &
    +2 \ln^2\frac{\bmu A^{12}e^{-\frac12}}{16\pi T}
    - \frac{7}{2}\ln^2(2)
    - 2(\ln\zeta_2)'^{2}
    +\frac{2\zeta_2''}{\zeta_2}
    +\frac{1}{2}\text{Li}_2\Bigl(\frac{2}{3}\Bigr)
    +\frac{1}{4}\text{Li}_2\Bigl(\frac{3}{4}\Bigr)
    -\frac{1}{2}
    \nn &
    -\frac14 \ln(3) \ln\frac{4}{3}
      \underbrace{-0.438129\vphantom{X_{X_X}}}_{q<2k}
    + \underbrace{1.97382\vphantom{X_{X_X}}}_{q>2k}
    \bigg\}
    \,.
    \label{finalLTfermion_app}
\end{align}

\subsubsection{Gluon and ghost contribution}
\label{sec:gluon_lmt}

We start from Eqs.~\eqref{deflt} and~\eqref{retdiffgluonsimple}
\begin{align}
  Z^{(c)}_{\mathrm{g},\rmii{$L$-$T$},\rmii{B}}&=i\frac{g^2\CA}{(4\pi)^2}
    \int_Q\int_{0}^{\infty}\!{\rm d}k\frac{(1+\nB(q^0))}{q^3(q^--i\varepsilon)^2}\nB(k)
    \bigg[
      \frac{q_{\perp}^2}{q^2(Q^2+i\varepsilon q^0)}\bigg(-8kq(3q_0^2-q^2)
    \nn &
    + Q^2(12k^2-q^2+3q_0^2)\bigg(
        \ln\frac{q_0+q-2k+i\varepsilon}{q_0-q-2k+i\varepsilon}
      + \ln\frac{q_0-q+2k+i\varepsilon}{q_0+q+2k+i\varepsilon}\bigg)
    \nn &
    + 12kq_0Q^2\bigg(
      2 \ln\frac{q_0+q+i\varepsilon}{q_0-q+i\varepsilon}
      + \ln\frac{q_0-q+2k+i\varepsilon}{q_0+q+2k+i\varepsilon}
      - \ln\frac{q_0+q-2k+i\varepsilon}{q_0-q-2k+i\varepsilon}\bigg)
    \bigg)
    -\text{adv.}\bigg]
    \,.
\end{align}
As was done in Sec.~\ref{app:eucl}, we need to separate out the contributions from zero and non-zero $Q$ Matsubara modes. In this case, because of the presence of the $1/(q^- - i\varepsilon)^2$ factor, the $q^0$ integrand possesses poles in addition to those associated with the Matsubara frequencies, meaning that we cannot effectively trade the $q^0$ integral for a sum over Matsubara modes as was done before. However, by first shifting: $q^z\rightarrow q^z+\frac{x^0}{x^z}q^0$, we may eliminate the $q^0$ pole tied to the $1/(q^- - i\varepsilon)^2$ prefactor, before proceeding to carry out the $q^0$ integration using contour methods. Of course, one may worry that such a shift spoils the analytical structure of the retarded (advanced) functions in the upper (lower) $q^0$ plane. It has nevertheless been pointed out by Caron-Huot \cite{CaronHuot:2008ni}%
\footnote{%
See also \cite{Ghiglieri:2015zma,Ghiglieri:2020dpq}.
}
that as long as $X$ is lightlike or spacelike, one may always work in a frame where $\frac{|x^0|}{|x^z|}\leq 1$, in which case the analytical structure of the causal functions is not altered.%
\footnote{%
While the methods introduced in~\cite{CaronHuot:2008ni} are strictly speaking only defined for $\frac{|x^0|}{|x^z|}<1$,
they can be safely extended to null correlators as well, provided the zero mode contribution is free of collinear singularities, which is the case here.
}
Thus, in this case where $X$ is lightlike, the appropriate shift is precisely: $q^z\rightarrow q^z+q^0$. Following this maneuver, we can safely carry out the integral $q^0$ and extract the $n=0$ contribution 
\begin{align}
Z^{(c)}_{\mathrm{g},\rmii{$L$-$T$},\rmii{B},n=0}&=\frac{g^2\CA T}{8\pi^2}
    \int_{\vec{q}}\int_{0}^{\infty}{\rm d}k\frac{q_{\perp}^2\nB(k)}{q^5(q^z +i\varepsilon)^2}\Bigg[4kq-(12k^2-q^2)\ln\bigg\vert\frac{2k-q}{q+2k}\bigg\vert\Bigg]
    \nn &=
    \frac{g^2\CA T^2}{32\pi^2}\left[
        -\frac{1}{\epsilon}+4
        -\ln\frac{\bmu^2 e^{\gammaE}}{4T^2}
        -\psi\Bigl(\frac{3}{2}\Bigr)
      \right]
    \,,
    \label{LmTzeromodediv}
\end{align}
which is needed for the matching calculation completed in Sec.~\ref{sec:matching:to:EQCD}.

For the contribution from the non-zero Matsubara modes, we rather persist in the Minkowskian signature so that we can employ the same strategy as in Appendix~\ref{app:fermion_lmt}. In more detail, for the HTL part, the $k$ integrand is given by $k\,\nB(k)$.
Therefore, given that
\begin{equation}
    \int_{0}^{\infty}\!{\rm d}k\,k\,\nB(k)=2\int_{0}^{\infty}{\rm d}k\,k\,\nF (k)=\frac{\pi^2T^2}{6}
    \,,
\end{equation}
we may simply make the replacement $2\Nf\Tf\rightarrow \CA$ in Eq.~\eqref{zcexprLTHTLcutfinal}
to obtain the bosonic equivalent.
This yields 
\begin{align}
    Z_{\mathrm{g},\rmii{$L$-$T$},\rmii{B},\rmii{HTL}}^{(c)}&=
   \frac{g^2 \CA T^2}{48\pi^2}\bigg[
      \frac{1}{\epsilon^2}
    + \frac{2}{\epsilon} \ln\frac{\bmu e^{\gammaE+\frac12}}{4\pi T}
    + 2 \ln^2\frac{\bmu e^{\gammaE+\frac12}}{4\pi T}
    - 4\gamma _1
    + \frac{\pi ^2}{4}
    -2 \gammaE^2+\frac32
    \bigg]
  \nn &
    +\frac{g^2 \CA T^2}{24\pi^2}(\ln2-1)\bigg[
      \frac{1}{\epsilon}+2\ln\frac{\bmu e^{\gammaE+\frac12}}{4\pi T}\bigg]
    \,.
    \label{zcexprLTHTLDRfinalg}
\end{align}

For the vacuum part of the non-HTL contribution,
we have instead
\begin{align}
\label{zcexprlogssymvacsympg}
  Z_{\mathrm{g},\rmii{$L$-$T$},\rmii{B},\rmii{non-HTL}\,\rmi{vac}}^{(c)}&
  =\frac{g^2 \CA}{16\pi}\int_0^\infty\!{\rm d}k \, \nB(k) \int_0^\infty\frac{{\rm d}q^0}{2\pi}
    \int_{\vec{q}}\frac{q_\perp^2}{q^5}\bigg[\frac{1}{(q^0-q^z-i\varepsilon)^2}
    +\frac{1}{(q^0+q^z+i\varepsilon)^2}\bigg]\nn
    &\times
  \bigg\{\theta(q-2k)\bigg[\left (3(2k-q^0)^2-q^2\right)\theta(q+2k-q^0)\theta(q^0-q+2k)
  \nn &
  \hphantom{\times\bigg\{\theta(q-2k)}
  - 24 k q^0\theta(q-2k-q^0)\bigg]
   \nn &
  +\theta(2k-q)\left (3(2k-q^0)^2-q^2\right)\theta(2k+q-q^0)\theta(q^0-2k+q)
\bigg\}
    \,,\\[2mm]
\label{zcexprlogssymvacsympfinalg}
  Z_{\mathrm{g},\rmii{$L$-$T$},\rmii{B},\rmii{non-HTL}\,\rmi{vac}}^{(c)}&=
  \frac{g^2 \CA T^2}{48\pi^2}\bigg\{
      \frac{1}{\epsilon^2}
    + \frac{2}{\epsilon}\ln\frac{\bmu A^{12}e^{-\frac12}}{8\pi T}
    + 2 \ln^2\frac{\bmu A^{12}e^{-\frac12}}{8 \pi T}
    + \frac{1}{2}\ln^2(2)
    - 2(\ln\zeta_2)'^{2}
    \nn &\hphantom{{}=\frac{g^2 \CA T^2}{48\pi^2}\bigg\{}
    +\frac{2\zeta_2''}{\zeta_2}
    +\frac12\mbox{Li}_2\Bigr(\frac{2}{3}\Bigr)
    +\frac14\mbox{Li}_2\Bigr(\frac{3}{4}\Bigr)
    -\frac12
    -\frac14 \ln(3) \ln\frac{4}{3}
    \bigg\}
  \,,
\end{align}
where in the last step, the methods of Appendix~\ref{app:fermion_lmt} were used.
For the thermal part, we have to subtract the zero mode (cf.\ Eq.~\eqref{LmTzeromodediv}%
\footnote{%
No zero-mode subtraction was necessary in the 
HTL part, since the zero mode is scale-free there.}).
This yields
\begin{align}
  Z_{\mathrm{g},\rmii{$L$-$T$},\rmii{B},\rmii{non-HTL}\,\rmi{th},n\neq 0}^{(c)}&=\frac{g^2\CA}{4\pi}\int_0^\infty\!{\rm d}k \, \nB(k) \int_0^\infty\frac{{\rm d}q^0}{2\pi}
    \int_{0}^\infty \frac{{\rm d}q }{4\pi^2}\frac{1}{q}\bigg[\nB(q^0)-\frac{T}{q^0}\bigg]
    \nn &\times
    \bigg[
        \frac{2 q^0 \tanh^{-1}\left(\frac{2 q q^0}{q^2+q_0^2}\right)}{q^3}-\frac{4}{q^2}\bigg]
    \nn &\times
  \bigg\{\theta(q-2k)\bigg[\left(3(2k-q^0)^2-q^2\right)\theta(2k+q-q^0)
  \nn &\hphantom{\times\bigg\{\theta(q-2k)}
  -\left(3(2k+q^0)^2-q^2\right)\theta(q-2k-q^0)\bigg]
  \nn &
  + \theta(2k-q) \left(3(2k-q^0)^2-q^2\right)\theta(2k+q-q^0)
  )\theta(q^0+q-2k)\bigg\}
  \,.
  \label{zcexprlogssymthetas3Dthermalg}
\end{align}
Here, we first perform the $q$ integral analytically,
followed by the $q^0$ and $k$ integrals
numerically.
We find
\begin{equation}
  Z_{\mathrm{g},\rmii{$L$-$T$},\rmii{B},\rmii{non-HTL}\,\rmi{th}\,n\neq 0}^{(c)}=
  \frac{g^2 \CA T^2}{48\pi^2}\bigg[
    \underbrace{0.145277\vphantom{X_{X_X}}}_{q<2k}
    \underbrace{-5.13832\vphantom{X_{X_X}}}_{q>2k}
    \bigg]
  \,.
\end{equation}
Not including the $Q$ zero-mode contribution, the total is then
\begin{align}
  Z_{\mathrm{g},\rmii{$L$-$T$},\rmii{B},n\neq 0}^{(c)}&=\frac{g^2 \CA T^2}{48\pi^2}
    \bigg\{\frac{2}{\epsilon^2}
    + \frac{2}{\epsilon} \Bigl(
        \ln\frac{\bmu e^{\gammaE+\frac12}}{4\pi T}
      + \ln\frac{\bmu A^{12}e^{-\frac12}}{8\pi T}
      + \ln2-1
      \Bigr)
    \nn &
    + 2 \ln^2\frac{\bmu e^{\gammaE+\frac12}}{4\pi T}
    - 4 \gamma_1+\frac{\pi ^2}{4}
    - 2 \gammaE^2+\frac32
    + 2(\ln2-1)\bigg[
      2\ln\frac{\bmu e^{\gammaE+\frac12}}{4\pi T}\bigg]
 \nn &
    +2 \ln^2\frac{\bmu A^{12}e^{-\frac12}}{8\pi T}
    + \frac{1}{2}\ln^2(2)
    - 2(\ln\zeta_2)'^{2}
    + \frac{2\zeta_2''}{\zeta_2}
    + \frac12\mbox{Li}_2\Bigl(\frac{2}{3}\Bigr)
    + \frac14\text{Li}_2\Bigl(\frac{3}{4}\Bigr)
    - \frac12
   \nn &
    - \frac14 \ln(3) \ln\frac{4}{3}
    + \underbrace{0.145277\vphantom{X_{X_X}}}_{q<2k}
      \underbrace{-5.13832\vphantom{X_{X_X}}}_{q>2k}
    \bigg\}\,.
    \label{finalLTgluon_app}
\end{align}

\section{Determination of the QCD coupling}
\label{app:coupling}

\begin{table}[t]
    \centering
    \begin{tabular}{|r|l|c|l|}
    \hline
        \multicolumn{1}{|c|}{$T$} &
        \multicolumn{1}{|c|}{$g^2$} &
        \multicolumn{1}{|c|}{$\Nf$} &
        \multicolumn{1}{|c|}{$\displaystyle\frac{Z_{\mathrm{g}}^{\rmi{3d}}\big\vert^\text{merge\vphantom{X}}}{\gE^2 T}$}
        \\[8pt]\hline
        $250$~MeV & $3.725027$ & 3 &$\hphantom{-}0.085(37)(12)(2)$\\
        $500$~MeV & $2.763516$ & 3 &$-0.024(36)(14)(1)$\\
          $1$~GeV & $2.210169$ & 4 &$-0.026(32)(4)(3)$\\
        $100$~GeV & $1.066561$ & 5 &$-0.179(15)(5)(2)$\\
        \hline
    \end{tabular}
    \caption{The values for the QCD coupling squared and the number of light flavors 
    correspond to those used in the matching coefficients in the lattice ensembles of \cite{Moore:2020wvy,Ghiglieri:2021bom}, see~\cite{Moore:2019lgw,Moore:2021jwe}
    for details. Data for $Z_{\mathrm{g}}^{\rmi{3d}}\big\vert^\text{merge}$ comes
from~\cite{Ghiglieri:2021bom}. }
    \label{tab:coupling}
\end{table}

When adding  perturbative and lattice determinations we need to be consistent in our choice of the coupling
$g$. Indeed, we need to set it to the values
that correspond to those chosen in the EQCD matching coefficients in the lattice ensembles. 
The matching procedure 
and coupling determination for these ensembles is detailed in~\cite{Moore:2019lgw,Moore:2021jwe},
following~\cite{Kajantie:1997tt,Laine:2005ai}.
The resulting values of the coupling and the number of light flavors $\Nf$
are listed in Tab.~\ref{tab:coupling}, together with the data for $Z_{\mathrm{g}}^{\rmi{3d}}\big\vert^\text{merge}$
from~\cite{Ghiglieri:2021bom}.%
\footnote{In the notation and normalisation of~\cite{Ghiglieri:2021bom}, that quantity is 
$(\Zg^\rmi{3d}/\gE^2)_\text{full}$.
It is listed in Table~1 there; please refer to the latest arXiv
version, which fixes a numerical issue in previous versions.
}
The matching procedure  
corresponds to a renormalisation scale 
\begin{equation}
    \label{fixmuuv}
    \bmu_\rmii{UV}= 4\pi T  e^{-\gammaE}\hat{\mu}
    \,,
\end{equation}
where at one-loop level~\cite{Kajantie:1997tt}
\begin{equation}
    \label{defmuhat}
    \hat{\mu} =  \exp\frac{-\Nc + 4\ln(4)\Nf}{22 \Nc - 4\Nf}
    \,.
\end{equation}
Hence, we set $\bmu_\rmii{UV}$ to Eq.~\eqref{fixmuuv} when
numerically evaluating Eq.~\eqref{zcqmsqmfinal} in Sec.~\ref{sec:discussion}.

{\small

\cleardoublepage
}
\end{document}